\documentclass[aps,prb,twocolumn,superscriptaddress,showpacs,showkeys]{revtex4-2}

\let\revappendix\appendix
 
\usepackage{amsfonts,bm}
\usepackage{amssymb}
\usepackage{amsmath}
\usepackage{graphicx} 
\usepackage{caption, subcaption}
\usepackage{varwidth}
\usepackage{chngcntr}
\usepackage{comment}
\usepackage{bm}
\usepackage{tabularx}
\usepackage{color}
\newcolumntype{C}{>{\centering\arraybackslash}X}
\newcolumntype{M}[1]{>{\centering\arraybackslash}m{#1}}

\newcommand{\newc}{\newcommand}
\newc{\beq}{\begin{equation}}
\newc{\eeq}{\end{equation}}
\newc{\beqa}{\begin{eqnarray}}
\newc{\eeqa}{\end{eqnarray}}
\newc{\baln}{\begin{align}}
\newc{\ealn}{\end{align}}
\newc{\balna}{\begin{align*}}
\newc{\ealna}{\end{align*}}
\newc{\mbf}{\mathbf}
\newc{\dg}{\dagger}
\newc{\rgtaw}{\rightarrow}
\newc{\lftaw}{\leftarrow}
\newc{\lrgtaw}{\longrightarrow}
\newc{\llfttaw}{\longrightarrow}
\newc{\mbhe}{{\mbf{\hat{e}}}}
\newc{\dstl}{\displaystyle}
\newc{\jijab}[2]{{J_{i \alpha; j \beta}^{{#1}{#2}}}}
\newc{\jijabSp}[4]{{J_{i {{#1}}; j {{#2}}}^{{#3}{#4}}}}
\newc{\jijabB}[2]{{J_{i \alpha; j \beta}^{{#1}{#2}, B}}}
\newc{\jjiba}[2]{{J_{j \beta; i \alpha}^{{#1}{#2}}}}
\newc{\jtab}[2]{{{\tilde{J}}_{\alpha \beta}^{{#1}{#2}}}}
\newc{\jtba}[2]{{{\tilde{J}}_{\beta \alpha}^{{#1}{#2}}}}
\newc{\jtaa}[2]{{{\tilde{J}}_{\alpha \alpha}^{{#1}{#2}}}}
\newc{\sia}[1]{{S}_{i \alpha}^{#1}}
\newc{\sjb}[1]{{S}_{j \beta}^{#1}}
\newc{\mDijab}[2]{{{\mathcal{D}}_{i \alpha; j \beta}^{{#1}{#2}}}}
\newc{\mDijabSp}[4]{{{\mathcal{D}}_{i {{#1}}; j {{#2}}}^{{#3}{#4}}}}
\newc{\GmInvqab}[4]{{\left[ {{\mathbf{\Gamma}}^{-1}}({#1}) \right]}^{#4}_{{#2},{#3}}}
\newc{\bdg}[1]{b_{{#1}}^\dg}
\newc{\fdg}[1]{f_{{#1}}^\dg}
\newc{\hf}[2]{{{\hat{f}}_{#1}(#2)}}
\newc{\hfdg}[2]{{{\hat{f}^{\dg}}_{#1}(#2)}}
\newc{\hHt}[1]{{\hat{H}}_{{int}}({#1})}
\newc{\ia}{{i \alpha}}
\newc{\jb}{{j \beta}}
\newc{\sumiajb}{\sum_{\substack{ {i \alpha, ~j \beta} \\ {}}} }
\newc{\sumlims}[2]{\sum\limits_{{#1}}^{#2}}
\newc{\sumliajb}{\sumlims{i\alpha, j\beta}{}}
\newc{\sumlqab}{\sumlims{\mbf{q}, \alpha, \beta}{}}
\newc{\sumsbstck}[2]{{\sum_{\substack{ {{#1}} \\ {{#2}}}} }}
\newc{\mcl}{\mathcal}
\newc{\jtabp}[2]{{{\tilde{J}}_{\alpha \beta^{\prime}}^{{#1}{#2}}}}
\newc{\jtbpa}[2]{{{\tilde{J}}_{\beta^{\prime} \alpha}^{{#1}{#2}}}}
\newc{\nia}{{\hat{n}}_{\ia}}
\newc{\njb}{{\hat{n}}_{\jb}}

\begin{document}

\title{Spin wave interactions in the pyrochlore Heisenberg antiferromagnet with Dzyaloshinskii-Moriya interactions}
\author{V. V. Jyothis}
\email{jyothis.vv@niser.ac.in}
\affiliation{School of Physical Sciences, National Institute of Science Education and Research Bhubaneswar, Jatni, Odisha 752050, India}
\affiliation{Homi Bhabha National Institute, Training School Complex, Anushaktinagar, Mumbai 400094, India}
\author{Kallol Mondal}
\email{kmondal@agh.edu.pl (Current address)}
\affiliation{School of Physical Sciences, National Institute of Science Education and Research Bhubaneswar, Jatni, Odisha 752050, India}
\affiliation{Homi Bhabha National Institute, Training School Complex, Anushaktinagar, Mumbai 400094, India}
\affiliation{AGH University of Krakow, Faculty of Physics and Applied Computer Science, Aleja Mickiewicza 30, 30-059 Krakow, Poland   }
\author{Himanshu Mavani}
\email{hmavani2@huskers.unl.edu (Current address)}
\affiliation{School of Physical Sciences, National Institute of Science Education and Research Bhubaneswar, Jatni, Odisha 752050, India}
\affiliation{Homi Bhabha National Institute, Training School Complex, Anushaktinagar, Mumbai 400094, India}
\affiliation{Department of Physics and Astronomy and Nebraska Center for Materials and Nanoscience,
University of Nebraska, Lincoln, Nebraska 68588-0299, USA  }
\author{V. Ravi Chandra}
\email{ravi@niser.ac.in}
\affiliation{School of Physical Sciences, National Institute of Science Education and Research Bhubaneswar, Jatni, Odisha 752050, India}
\affiliation{Homi Bhabha National Institute, Training School Complex, Anushaktinagar, Mumbai 400094, India}

\begin{abstract}
We study the effect of magnon interactions on the spin wave spectra of the all-in-all-out phase of the pyrochlore nearest neighbour
antiferromagnet with a Dzyaloshinskii-Moriya interaction ($D$). The leading order corrections to spin wave energies indicate a significant 
renormalisation for commonly encountered strengths of the Dzyaloshinskii-Moriya term. For low values of $D$ we find a potential instability of the
phase itself, indicated by the renormalisation of magnon frequencies to negative values. 
We have also studied the renormalized spectra in the presence of magnetic fields along three high symmetry directions
of the lattice, namely the $[111]$, $[100]$ and $[110]$ directions. Generically, we find that for a fixed value of the Dzyaloshinskii-Moriya interaction
renormalized spectra for the lowest band decrease with an increasing strength of the field. We have also analyzed the limits of the
two magnon continuum and probed the possibility of magnon decay. For a range of $D$ and the field strength we identify possible parameter
regimes where the decay of the higher bands of the system are kinematically allowed.

\end{abstract}

\maketitle

\section{Introduction}
\label{intro_section}
The study of the effects of interactions between magnons
is nearly as old as the description of magnons as the lowest excitations
in ordered magnets. After early descriptions of spin waves
as elementary excitations of ordered ferromagnets \cite{Bloch_SW_1930} and antiferromagnets
\cite{Anderson_SW_1952, Kubo_1952_SW},
detailed analyses of magnon interactions in both ferromagnets
and antiferromagnets were presented 
\cite{Dyson_SW_FM_interactions_1, Dyson_SW_FM_interactions_2, Oguchi_SW_interactions_1960}.
Most of these early efforts were focussed on the effect of spin wave interactions in spin models
with collinear magnetic order, usually for lattices with cubic symmetry and purely Heisenberg 
interactions. For such cases, at least at low temperatures, it was concluded
that the phases could be described fairly accurately using non-interacting magnons. For instance,
the results for relevant physical quantities like magnetization, specific heat, susceptibility etc.
do not deviate from those of linear spin wave theory (LSWT) beyond a few percent if the leading
corrections due to magnon interactions are taken into account 
\cite{Dyson_SW_FM_interactions_1, Dyson_SW_FM_interactions_2, Oguchi_SW_interactions_1960}.\\

Many of the ordered magnetic materials and the model Hamiltonians of interest currently 
do not possess the above mentioned characteristics that render spin wave interactions irrelevant.
In the present day, there is a lot of interest in the kind of magnetic order which is not a
standalone phase but rather competes with other ordered phases or spin liquid states.
Furthermore, the occurence of collinear order (moments pointing along or opposite to a single direction) 
is more likely to be a feature of spin models
with a bipartite lattice structure and short range interactions.
Many of the contemporary magnetic Hamiltonians that 
have long ranged magnetic order have non-collinear ordering patterns, either because of multiple
coupling constants or geometrical frustration or both.
Finally, the analysis of many measurements requires considerations of excitations of spin models
away from the lowest temperatures. Magnon interactions can in principle play a significant role in all such circumstances.
The effect of spin wave interactions in such systems is far from benign and can lead to substantial
spectral renormalisation and additional effects like spontaneous decay of magnons 
\cite{SW_in_triangular_AFM_Chernyshev_Zhitomirsky_prb_2009, magnon_decay_RMP_2013}. \\

Most of the above reasons to study spin wave interactions can arise simultaneously
in spin models on the pyrochlore lattice. It is one of the most
widely studied platforms to probe the effect of geometrical frustration
on magnetic order. The $A_2 B_2 O_7$ class of materials
provide an array of possible material realizations \cite{gardner_gingras_greedan_rmp}.
Furthermore, this lattice also proved to be the source of one of the first
topological magnonic band structures investigated \cite{onose_et_al_science_2010},
where a non-vanishing thermal hall conductivity was explained using a finite 
Berry curvature of the spin wave bands. In this context, it is important to note
that while understanding of spin wave interactions is relevant
and important in its own right, interest in the same has also witnessed a resurgence
recently because of the interest studying topological 
magnonic band structures \cite{McClarty_Top_Mag_review, Malki_top_mag_review_2020, kondo_akagi_katsura_2020_ptep}.
Much of the analysis of topological magnons relies on the magnon band structure
at the non-interacting level. However, the emergence of topologically non-trivial spin wave band structures
often involves non-collinear magnetic order and/or geometrically frustrated lattices,
circumstances in which deviations from linear spin wave theory are frequently found.
As a result there have been several recent attempts to ascertain the effect of interactions on
the predictions of topological magnonic theories, mostly for two dimensional Hamiltonians
\cite{dampped_TM_in_Kagome_Chernyshev_Maximov_prl_2016, Pershoguba_et_PRX_2018,
mook_et_al_interaction_stabilised_TMI_2022, Liu_et_al_SW_interactions_MnBi2Te4_prb_2023, honeycomb_damping_chen_etal_prb_2023,
Habel_et_al_PRB_2023, Sun_Bhowmick_Yang_Sengupta_PRB_2023, sourounis2024PRB, Li_Luo_Chang_T_induced_Chern_insulator_PRB_2023}. \\
The lattice structure of the pyrochlore allows for anisotropic interactions like the Dzyaloshinskii-Moriya interaction (DMI)
and local spin anisotropies which make it a very suitable system for non-collinear low temperature phases and thence for
the study of magnon interactions.
Existing work on the effect of spin wave interactions in this lattice have focussed on different aspects of the ferromagnetic phase 
\cite{Rau_Moessner_McClarty_NLSWT_PRB_2019, Hickey_et_al_2025_PRB}.
In this work we initiate a study of the effect of spin wave interactions in a spin model on
the pyrochlore lattice with antiferromagnetic Heisenberg exchange and Dzyaloshinskii-Moriya interactions.
We study the all-in-all-out (AIAO) phase of this lattice and the effect of magnon interactions 
on spin wave spectra of this phase.
Several aspects of the magnon spectra of this phase both in the bulk and the thin film limits have been
studied recently
\cite{Li_Chen_spin_1_pyrochlore_2018, Laurell_Fiete_2017_prl, Li_Chen_spin_1_pyrochlore_2018, 
Jian_Nie_Weyl_2018, 
hwang_et_al_prl_2020, Jyothis_et_al_JPCM_2024}. The frequent occurrence of this
phase in the pyrochlore lattice and the nature of the phase itself make it an important
target for the investigation of spin wave interactions.
\begin{figure*}[t]
\centering
\includegraphics[width=1.0\textwidth]{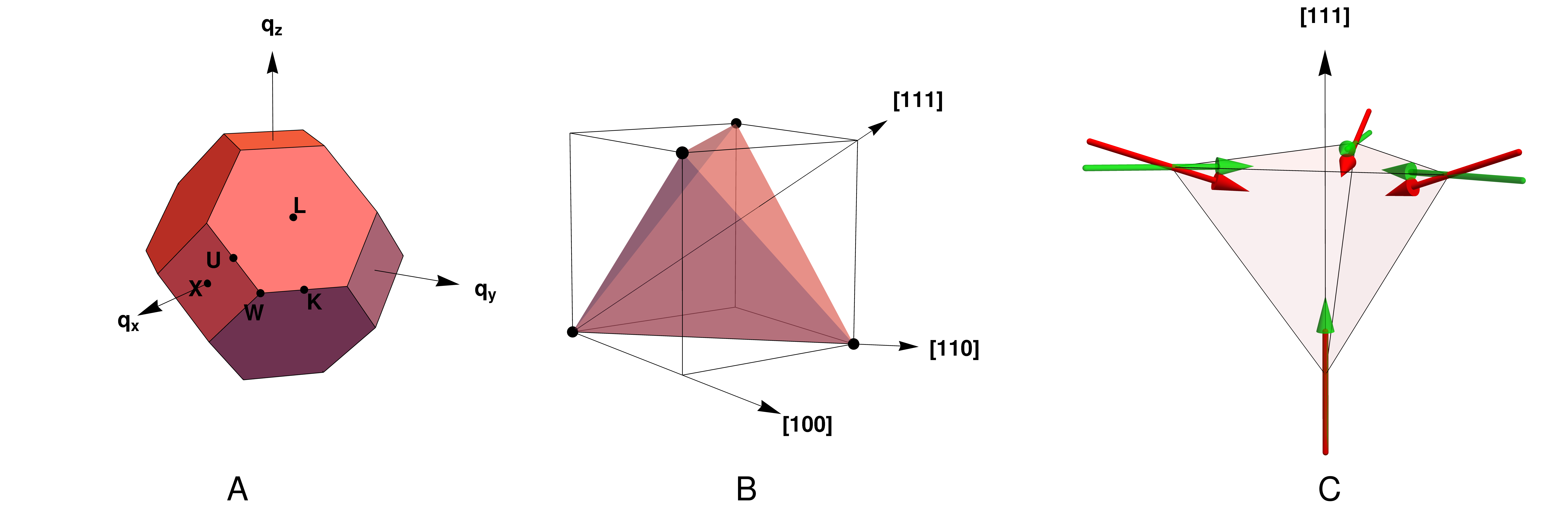}
\caption{{\small{{
{{\bf{A}}}. The Brillouin zone of the pyrochlore lattice with the locations of the high symmetry points used in this work.
{{\bf{B}}}. The three high symmetry directions along which magnetic fields have been considered.
        {{\bf{C}}}. The classical ground states in the absence and the presence of a magnetic field.
        The red colored arrows are spin directions in the all-in-all-out phase which is the ground state in the absence of a field for a finite $D(=0.1J \textrm{~here})$.
        The green colored arrows are after a finite field $(B=2.0J)$ is applied. The field is along $[111]$ direction in this example.
}}}}
\label{Fig_BZ_Bdirs_GS}
\end{figure*}
\section{Model Hamiltonian and formalism}
\label{model_and_formalism_section}
The model Hamiltonian we study has the form: 

\begin{align}
H &= J \sum_{<i\alpha, j\beta>} \mathbf{S}_{i \alpha} \cdot \mathbf{S}_{j \beta} + \sum_{<i\alpha, j\beta>} {\mbf{D}}_{i\alpha, j\beta} \cdot (\mathbf{S}_{i \alpha} \times \mathbf{S}_{j \beta}) \nonumber \\
        & ~~~~~~- \sum_{i\alpha} \mbf{B} \cdot \mathbf{S}_{i \alpha}  \label{basic_H_test}  
\end{align}

Here $\mathbf{S}_{i \alpha}$ are spins on the sites of the pyrochlore lattice and the first subscript denotes the 
Bravais lattice index and the Greek index denotes one of the four sub lattices. 
$J$ and ${\mbf{D}}_{i\alpha, j\beta}$ are 
the coefficients of the Heisenberg exchange and Dzyaloshinskii-Moriya interactions \cite{Dzyaloshinsky_1958, Moriya_DMI_prl, Moriya_DMI_Phys_Rev}
respectively. $\mbf{B}$ is a static magnetic
field and in our analysis we consider fields pointing along the three commonly used high symmetry directions of the lattice
namely, $[111], [100]$ and the$ [110]$ directions. The parameters $D$ and $B$ wherever they appear as numerical values are quoted in units of $J$ which
has the dimensions of energy. We develop the formalism below for a general spin quantum number $S$ and present most
numerical results for $S=1$ and make comments about $S=1/2$ at appropriate places.\\

In the absence of a magnetic field and the DMI term, 
the Hamiltonian with only a nearest neighbour exchange term
has a macroscopically degenerate classical ground state and is a well known example of a classical
spin liquid \cite{moessner_and_chalker_PRL_1998, moessner_and_chalker_PRB_1998}. One set of
states of this degenerate manifold are the ordered states of the all-in-all-out kind.
A direct $(D>0)$ DMI term in the Hamiltonian selects this particular ordered state out
of the degenerate manifold as the classical ground state, as noted in early work \cite{elhajal_et_al_pyrochlore_dm} and which can also be seen from a Luttinger-Tisza analysis \cite{luttinger_and_tisza_1946}. 
Thus the classical ground state is one which has the translation invariance of the underlying FCC lattice. \\ 

Measurements and characterization of pyrochlore materials often involve applying a magnetic field
along high symmetry directions of the lattice. 
We analyzed the renormalized spectra for fields of various strengths applied along three of the commonly
utilized high symmetry directions of this lattice, namely, the $[111], [100]$ and $[110]$ directions shown 
in Fig.\ref{Fig_BZ_Bdirs_GS}B.
In the presence of a magnetic field, the classical ground state naturally differs from
the AIAO state at zero field. Using numerical minimisation within the space
of states in which spins belonging to a particular sublattice have the same orientation, 
we find the classical ground states for finite fields. 
We study fields pointing along one of the three high symmetry directions mentioned above. 
As expected on general grounds we find that for finite fields the moments
tilt towards the field direction by a magnitude depending on the strength of the field. 
While the qualitative effect of the moments canting towards the field is similar for fields in all the
three high symmetry directions, the details of how the moments tilt can depend on the direction
of the field. For the $[111]$ direction one of the moments is already aligned with the field
and the other three moments tilt by the same angle towards the field as shown in Fig.\ref{Fig_BZ_Bdirs_GS}C.
The canting happens in this case in such a manner that each spin rotates towards the field
in a plane defined by its initial direction and the field direction.
For the fields in the $[100]$ direction too, this property holds. However, the magnitude of tilt is not the
same for all the spins, one pair of sublattices have one tilt value and another pair have a different
tilt value. Finally, for the $[110]$ direction, the tilt towards the field is not in general in the plane
defined by the initial moment direction and the field direction.
For the field strengths that we have investigated, the tilting effect of a magnetic field on the 
classical ground state is small, as depicted
for a representative case in Fig.\ref{Fig_BZ_Bdirs_GS}C. 
It shows the moments of the tetrahedral unit in the classical ground state without and with a magnetic field
(of strength $B=2.0J$) for $D=0.1J$. Even though we have only analysed field strengths far from saturation
it might be of interest to the reader to have an estimate of the saturation fields. For the case of the field in the $[111]$ direction, the energy function to be minimized can be written in an analytical form
as a function of the tilt angle. In Appendix \ref{canting_angle}, we present that function, and under
certain stated assumptions, present how the canting angle changes with the field up to saturation. We 
emphasise again however, that the data presented in the rest of main text is for fields
far away from saturation.\\ 
Before we discuss our results for the renormalized spectra of the AIAO phase,
we present below the relevant aspects of the formalism of LSWT and
magnon interactions in general terms, in the notation that we use throughout the text.\\
We assume the spin model to be defined on a lattice which is a Bravais lattice with a number of sublattices. 
The classical ground state phases we consider are such that each spin belonging to a sublattice has a fixed orientation. Using a suitable rotation of axes we orient the local "z-axis" 
(denoted using the superscript $3$ below) along the direction of the moment at each sublattice.
The Hamiltonian $H$ of Eq. \ref{basic_H_test} can be written as:
\begin{multline} \label{sw_ham_ladder_ops}
  H = \sum_{\substack{ {i \alpha, ~j \beta} \\ {}}} \bigg [ \jijab{3}{3} \sia{3} \sjb{3} \\ 
        +  \jijab{+}{+} \sia{+} \sjb{+} + \jijab{-}{-} \sia{-} \sjb{-}   \\
        +   \jijab{+}{-} \sia{+} \sjb{-} + \jijab{-}{+} \sia{-} \sjb{+}  \\
  + \jijab{+}{3} \sia{+} \sjb{3} + \jijab{3}{+} \sia{3} \sjb{+} \\ 
  + \jijab{-}{3} \sia{-} \sjb{3} + \jijab{3}{-} \sia{3} \sjb{-}  \bigg ] \\
  - \sum_{i\alpha} B_{\alpha}^3  \sia{3} - \frac{1}{2} \sum_{i\alpha} \left ( B_{\alpha}^{-} \sia{+} + B_{\alpha}^{+} \sia{-} \right ) 
\end{multline}

The sums over $i \alpha, ~j \beta$ above are unrestricted and run over all sites of the lattice. 
All coupling constants of the form $\jijab{\left [ \right ]}{\left [ \right]}$
have the translation invariance of the underlying Bravais lattice and periodic boundary conditions are 
assumed when finite lattices are studied. We define the local axes unit vectors for every sublattice $\alpha:$ 
$\dstl{\left ({\mbhe_{\alpha}}^1, ~\mbhe_{\alpha}^2, ~\mbhe_{\alpha}^3 \right )}$, with the moment pointing
along $\mbhe_{\alpha}^3$ in the classical ground state. 
If the orientation of the sublattice $\alpha$ is given by $(\theta_\alpha, \phi_\alpha)$ 
point on the unit sphere then we choose $\dstl{\left ({\mbhe_{\alpha}}^1, ~\mbhe_{\alpha}^2, ~\mbhe_{\alpha}^3 \right )} = \left ({\hat{\theta}}_\alpha, ~{\hat{\phi}}_\alpha, ~{\hat{r}}_\alpha \right )$.
$\dstl{\left ({\mbhe_{\alpha}}^1, ~\mbhe_{\alpha}^2, ~\mbhe_{\alpha}^3 \right )}$ define a local right handed coordinate system for each sublattice $\alpha$. 
So we have $\sia{m} = \mathbf{S}_{i \alpha} \cdot {\mbhe_{\alpha}}^m$ and $\sia{\pm} = \sia{1}\pm\sia{2}$
and thus $\sia{m}$ follow the standard spin commutation relations.
In the last set of terms in Eq. \ref{sw_ham_ladder_ops} the
scalar product $\mbf{B} \cdot \mbf{S}$ has been written using the components of ${B}_{\alpha}^{m}$ and $\sia{m}$ along the local axes 
at the location of each spin 
with $B_{\alpha}^{\pm} \equiv (B_{\alpha}^1 \pm i B_{\alpha}^2)$. As indicated above ${B}_{\alpha}^{3},\sia{3}$ are 
the components along the direction of the spin in the classical ground state.\\

We provide one example to make clear for the reader the relation between the terms 
in Eq. \ref{basic_H_test} and Eq. \ref{sw_ham_ladder_ops}. We consider for illustration, the
terms in Eq. \ref{basic_H_test} with the sublattice indices $\alpha=1, \beta=2$.
$\dstl{\left ({\mbhe_{1}}^1, ~\mbhe_{1}^2, ~\mbhe_{1}^3 \right )}$ are the  
orthonormal unit vectors of the local right handed coordinate system associated with sublattice-$1$. 
The vector operator for the spin at sublattice-$1$ can be written as:
$\dstl{ {\mbf{S}}_{i 1} = \sum_{m}  S_{i 1}^{m} ~\mbhe_{1}^m}$. The notation for $\dstl{{\mbf{S}}_{i 2}}$
is defined analogously. We note that $ \mbhe_{\alpha}^m  $ do not have a Bravais lattice site index, 
consistent with the fact that the studied classical ground state has the translation invariance of the 
underlying Bravais lattice. We re-express the exchange and DMI terms corresponding to this pair,
as follows:
\begin{multline} \label{sub_1_2_JDM_term}
\frac{1}{2} \left [ J \mathbf{S}_{i 1} \cdot \mathbf{S}_{j 2} + {\mbf{D}}_{i1, j2} \cdot (\mathbf{S}_{i 1} \times \mathbf{S}_{j 2}) \right ] = \\
\frac{1}{2} \left [   \sum_{\substack{{m, ~n} {}}} \left [ J \mbhe_{1}^m \cdot \mbhe_{2}^n + {\mbf{D}}_{i1, j2} \cdot \left ( \mbhe_{1}^m \times \mbhe_{2}^n \right) \right ] S_{i 1}^{m} ~  S_{j 2}^{n} \right ] =  \\
\sum_{\substack{{m, ~n} {}}} \mathcal{D}_{i 1; j 2}^{mn} ~ S_{i 1}^{m} ~  S_{j 2}^{n}
\end{multline}
The factor of $\dstl{\frac{1}{2}}$ has been introduced to account for the fact that the sums in 
Eq. \ref{sw_ham_ladder_ops} are unrestricted. In writing the above we have assumed that 
${(i1, j2)}$ are nearest neighbours. Given the unrestricted nature of the sums in 
Eq. \ref{sw_ham_ladder_ops}, if ${(i\alpha, j\beta)}$ are not a nearest neighbour pair, 
$\mDijab{m}{n}$ are zero for our Hamiltonian. 
We can now rewrite all the terms in the above expression in terms of 
$\sia{3}, \sia{\pm} = \sia{1} \pm i \sia{2}$. Doing so and collecting the coefficients
appropriately yields Eq. \ref{sw_ham_ladder_ops}. For instance, $\jijabSp{1}{2}{+}{+}$ is given by,
\begin{align}
\jijabSp{1}{2}{+}{+} = \frac{1}{4} \left[\mDijabSp{1}{2}{1}{1} - i\mDijabSp{1}{2}{1}{2} - i\mDijabSp{1}{2}{2}{1} - \mDijabSp{1}{2}{2}{2}  \right]  \label{jijab_pp_exp}
\end{align} 

Eqs. \ref{sub_1_2_JDM_term} and \ref{jijab_pp_exp} (and analogous ones for other sublattice pairs) 
establish the connection between the bare coupling 
constants, the classical ground state about which spin wave excitations are calculated, and the coupling constants in Eq. \ref{sw_ham_ladder_ops}. Though we chose our nearest neighbour example to illustrate the 
relation, the notation of course is general and can incorporate broadly defined Hamiltonians with terms 
that are quadratic in spin operators (including generic single site anisotropy terms).
The local axes conventions that we have used and the detailed expressions of all the coefficients 
$\jijab{\left [ \right ]}{\left [ \right]}$ in terms of $\mDijab{m}{n}$ following the same procedure
outlined above can be found in the Appendix of earlier work by two of us \cite{Jyothis_et_al_JPCM_2024},
where our notation was introduced. \\
\begin{figure*}[t]
\centering
\includegraphics[width=1.0\textwidth]{Figure_2_PRBPA.png}
\caption{{\small{{
A representative selection of linear spin wave theory (LSWT) spectra.
{{\bf{(A, B, C, D):}}}
Magnon bands of the LSWT Hamiltonian for $D=0.1J, 0.2J, 0.3J, 0.4J$ and $B\approx0 (=0.005J)$.
{{\bf{(E, F, G, H):}}} Magnon bands of the LSWT Hamiltonian for $D=0.3J$ and $B = 0.5J, 1.0J, 2.0J, 3.0J$ along the $[111]$ direction. The interacting spin wave analysis discussed in the rest of this work study the
leading order corrections to LSWT results for $D\in (0.1J, 0.2J, 0.3J, 0.4J, 0.5J)$ and $B\in (0.5J, 1.0J, 2.0J, 3.0J)$ along the $[111], [100]$, and $[110]$ directions.
}}}}
\label{Fig_LSWT_spectra_panel}
\end{figure*}
We use the Holstein-Primakoff (HP) \cite{holstein_primakoff_original_paper_1940} bosonic represention for the spin operators $\left(\sia{3} = S - b_{i\alpha}^\dg b_{i\alpha},~ \sia{+} = \sqrt{2S - b_{i\alpha}^\dg b_{i\alpha}}~b_{i\alpha}\right)$. The Hamiltonian Eq. \ref{sw_ham_ladder_ops} expressed in terms of $b_{i\alpha}$, retaining the leading order terms involving interactions among (HP) bosons have the structure:

\begin{widetext}
\beqa
& H \approx H_{SW} = H_{LSWT} + H_{int}, ~~H_{LSWT} =  ~\mcl{E}_{GS}^{LSWT} + {\sumlims{{\mbf{q}},\alpha}{}}^{} \left[ \epsilon_{\mbf{q} \alpha}~ f_{\mbf{q}\alpha}^\dg {f}_{\mbf{q}\alpha} \right ] \hspace{0.0cm} \label{H_SW_main}\\
& H_{int} =  H_{int}^{33} +  H_{int}^{++} +  H_{int}^{--} +  H_{int}^{+-} +  H_{int}^{-+} + 
H_{int}^{3+} +  H_{int}^{+3} +  H_{int}^{3-} +  H_{int}^{-3}, \textrm{where} \hspace{0.0cm} \label{H_int_Eq}  \\
& H_{int}^{33}  = \frac{1}{N} {{\sumsbstck{\mbf{\{q_i\}, Q}}{\alpha, \beta}}} \jtab{3}{3}(\mbf{Q}) \bdg{\mbf{q_1} \alpha} b_{\mbf{q_1} + \mbf{Q} \alpha} \bdg{\mbf{q_3} \beta} b_{\mbf{q_3}-\mbf{Q} \beta} \nonumber \\ 
& H_{int}^{++} = - \frac{1}{2N} {{\sumsbstck{\mbf{\{q_i\}}}{\alpha, \beta}}} \jtab{+}{+}(\mbf{q_2}) b_{\mbf{q_2} \alpha} \bdg{\mbf{q_2+q_3+q_4} \beta} b_{\mbf{q_3} \beta} b_{\mbf{q_4} \beta} - \frac{1}{2N} {{\sumsbstck{\mbf{\{q_i\}}}{\alpha, \beta}}}  \jtab{+}{+}(\mbf{-q_4})  \bdg{\mbf{q_2 + q_3 + q_4} \alpha} b_{\mbf{q_2} \alpha} b_{\mbf{q_3} \alpha} b_{\mbf{q_4} \beta} \hspace{0.0cm} \nonumber \\ 
& H_{int}^{+-}  =  - \frac{1}{2N} {{\sumsbstck{\mbf{\{q_i\}}}{\alpha, \beta}}} \jtab{+}{-}(\mbf{q_2 + q_3 - q_4}) b_{\mbf{q_2+q_3-q_4} \alpha} \bdg{\mbf{q_2} \beta} \bdg{\mbf{q_3} \beta} b_{\mbf{q_4} \beta} - \frac{1}{2N} {{\sumsbstck{\mbf{\{q_i\}}}{\alpha, \beta}}} \jtab{+}{-}(\mbf{q_2+q_3-q_4}) \bdg{\mbf{q_4} \alpha} b_{\mbf{q_2} \alpha} b_{\mbf{q_3} \alpha} \bdg{\mbf{q_2+q_3-q_4} \beta} \nonumber \\
& H_{int}^{3+} = - \sqrt{\frac{2S}{N}} {{\sumsbstck{\mbf{\{q_i\}}}{\alpha, \beta}}} \jtab{3}{+}(\mbf{-q_3}) \bdg{\mbf{q_2 + q_3} \alpha} b_{\mbf{q_2} \alpha} b_{\mbf{q_3} \beta} - \sqrt{\frac{S}{8N}} {{\sumsbstck{\mbf{\{q_i\}}}{\alpha, \beta}}} \jtab{3}{+}(\mbf{0}) \bdg{\mbf{q_2 + q_3} \beta} b_{\mbf{q_2} \beta} b_{\mbf{q_3} \beta} \hspace{0.0cm} \nonumber \\
& H_{int}^{+3} = - \sqrt{\frac{2S}{N}} {{\sumsbstck{\mbf{\{q_i\}}}{\alpha, \beta}}}  \jtab{+}{3}(\mbf{q_2-q_3}) b_{\mbf{q_2 - q_3} \alpha} \bdg{\mbf{q_2} \beta} b_{\mbf{q_3} \beta} - \sqrt{\frac{S}{8N}} {{\sumsbstck{\mbf{\{q_i\}}}{\alpha, \beta}}} \jtab{+}{3}(\mbf{0}) \bdg{\mbf{q_2 + q_3} \alpha} b_{\mbf{q_2} \alpha} b_{\mbf{q_3} \alpha}\hspace{0.0cm}  \nonumber
\eeqa
\end{widetext}
where the other terms $(H_{int}^{--},~ H_{int}^{-+},~H_{int}^{3-},~H_{int}^{-3})$ in Eq. \ref{H_int_Eq} are 
related to the terms written above because of the Hermiticity of the Hamiltonian. 
In the terms that we have written in Eq. \ref{H_int_Eq}, though the magnetic field is not seen 
explicitly it is included:the LSWT spectra $\epsilon_{\mbf{q} \alpha}$ depend on it explicitly and 
the coupling constants depend on it implicitly since the field plays a role in determining the classical 
ground state and the various coupling constants (see the discussion leading to Eqs. \ref{sub_1_2_JDM_term} 
and \ref{jijab_pp_exp}).\\

The non-interacting part of the Hamiltonian $H_{LSWT}$ is written in terms of new bosonic operators $\{ {f}_{\mbf{q}\alpha}\}$ 
$( \left[ {f}_{\mbf{q}\alpha}, \fdg{\mbf{q^{\prime} }\alpha^{\prime}}\right] = \delta_{q q^{\prime}} \delta_{\alpha \alpha^{\prime}} )$
and $\epsilon_{ \mbf{q} \alpha}$ are the spin wave energies at the level of non-interacting magnons. 
We denote by $\mbf{\Gamma_q}$, the linear transformation relating the old and the new bosonic operators, 
$[\mbf{f_q}~~~~~\mbf{\fdg{-q}}]^T = \mbf{\Gamma_q} [\mbf{b_q}~~~~~\mbf{\bdg{-q}}]^T$. 
For the construction of $\mbf{\Gamma_q}$ we follow a well established procedure 
\cite{colpa_diagonalisation_1978}. In the presence of a finite magnetic field we determine the 
classical ground state, necessary to proceed with the spin wave analysis, 
using numerical optimization. The evaluation of $\epsilon_{q\alpha}$ includes the
effect of a finite magnetic field term. We note that the field term also results in terms which are linear
in the transverse spin components but such terms (in conjunction with similar terms from the exchange terms) vanish for a valid extremum of the classical energy function.\\
$H_{int}$ is the magnon-magnon interaction part of the Hamiltonian, 
obtained by keeping the next order terms beyond those required for linear spin wave theory using
 Holstein-Primakoff bosons. We retain terms till $S^0$ in writing $H_{int}$ and consider leading 
order effects of nonlinear spin wave theory (NLSWT).\\ 

We consider the time ordered Green's function at zero temperature:
\beq \label{G_ab_first}
i {\bf{G}}_{\alpha \beta}(q, t_1-t_2) = \frac{\langle \mbf{\Psi}_{GS} \vert \mathcal{T} \left( {f}_{q \alpha}(t_1)  \fdg{q \beta}(t_2) \right ) \vert \mbf{\Psi}_{GS}  \rangle}{ \langle \mbf{\Psi}_{GS} \vert \mbf{\Psi}_{GS} \rangle} 
\eeq

Here $\vert \mbf{\Psi}_{GS} \rangle$ is the ground state of $H_{SW}$ in Eq. \ref{H_SW_main} and 
${f}_{q \alpha}(t_1)$ and $\fdg{q \beta}(t_2)$ are operators 
in the Heisenberg representation. $\mathcal{T}$ is the time-ordering operator which orders operators from latest to earliest times and $ i \equiv \sqrt{-1}$. 
Following the standard formulation for such Green's functions \cite{Fetter_and_Walecka} we have:

\begin{align} \label{G_ab_int}
i {\bf{G}}_{\alpha \beta}(q, t_1-t_2) = \hspace{4.5cm}  & \nonumber \\
        \frac{{}_f{\langle} \mbf{0} \vert \mathcal{T} \left [ \hf{q \alpha}{t_1} \hfdg{q \beta}{t_2} \exp \left (- i  \int_{-\infty}^{\infty} \hHt{t} dt \right ) \right ] \vert \mbf{0}  {\rangle}_f }{{}_f{\langle} \mbf{0} \vert \mathcal{T} \exp \left (- i  \int_{-\infty}^{\infty} \hHt{t} dt \right )  \vert \mbf{0} {\rangle}_f } & 
\end{align}

The $"~\hat{}~"$ above the operators indicates the interaction picture with the non-interacting Hamiltonian being $H_{LSWT}$. 
$H_{int}$ has already been defined in Eq. \ref{H_int_Eq}.  $\vert \mbf{0} {\rangle}_f$ is the vacuum state of the $f$-bosons $(f_{\mbf{q}\alpha} \vert \mbf{0} {\rangle}_f =0, \forall \{q, \alpha\})$
or equivalently, the ground state of $H_{LSWT}$. 
$ {\bf{G}}_{\alpha \beta}(q, \omega) = \int {\bf{G}}_{\alpha \beta}(q, t) \exp(i \omega t) dt $ satisfies the Dyson equation:
\begin{align} \label{G_ab_dyson}
   {\mbf{G}}_{\alpha \beta}(q, \omega) = \left[ \frac{1}{{\mbf{G}}_{0}^{-1}(q, \omega) -  \mbf{\Sigma} (q, \omega)} \right ]_{\alpha \beta}, \textrm{where,~} & \\
        [{\mbf{G}}_{0}]_{\alpha \beta} (q, \omega) = \left [ \frac{{1}}{\omega - \epsilon_{q \alpha} + {i} \eta }\right ] \delta_{\alpha \beta} \hspace{2.0cm} & \nonumber 
\end{align}

where $ \mbf{\Sigma} (q, \omega) $ is the irreducible self energy. The poles of the Green's function $ {\bf{G}}_{\alpha \beta}(q, \omega) $ determine the renormalized spectra
and their widths. In this work we calculate the leading order term of the irreducible self energy $ \mbf{\Sigma} (q, \omega) $ and evaluate the poles of the Green's function to obtain the renormalised spectra.\\

We note here the $H_{int}$ as written in Eq. \ref{H_int_Eq} is not normal ordered when expressed 
in terms of the $f$-bosons, in terms of which the vacuum of the non-interacting Hamiltonian is defined.  
In order to evaluate the first term in self energy using the application of Wick's theorem, we
use the prescribed method of normal ordering the operators of the interaction Hamiltonian which appear 
with the same time label \cite{Wicks_Theorem_original_paper}, before applying the contractions.\\  

Though the procedure is known, we provide here one illustrative example of how the contributions
to the leading order self energy are evaluated for our case, in our notation. \\
Consider one of the terms of the interaction Hamiltonian, namely $H_{int}^{33}$, which when 
written in terms of the $f-$bosons has the following form:
{\small{{
\begin{align}
H_{int}^{33} & = \frac{1}{N} {{\sumsbstck{\mbf{\{q_i\}, Q}}{\alpha, \beta}}} \jtab{3}{3}(\mbf{Q}) \bdg{\mbf{q_1} \alpha} b_{\mbf{q_1} + \mbf{Q} \alpha} \bdg{\mbf{q_3} \beta} b_{\mbf{q_3}-\mbf{Q} \beta} \hspace{1.0cm}\nonumber \\
& =\frac{1}{N}{{\sumsbstck{\mbf{\{q_i\}, Q}}{\alpha, \beta, \{\beta_i\}}}} \jtab{3}{3}(\mbf{Q}) \nonumber \\
\times & \left[\GmInvqab{\mbf{q}_1}{\alpha}{4+\beta_1}{*} f_{ \mbf{-q}_{1} {\beta}_1}+\GmInvqab{\mbf{q}_1}{\alpha}{\beta_1}{*} \fdg{\mbf{q}_1 \beta_1}\right] \nonumber \\
\times & \left[\GmInvqab{\mbf{\tilde{q}_1}}{\alpha}{\beta_2}{} f_{\mbf{\tilde{q}_1},\beta_2}+\GmInvqab{\mbf{{\tilde{q}}_1}}{\alpha}{4+\beta_2}{} \fdg{\mbf{-\tilde{q}_1}, \beta_2}\right] \nonumber \\
\times & \left[\GmInvqab{\mbf{q}_3}{\beta}{4+\beta_3}{*} f_{ \mbf{-q}_{3} {\beta}_3}+\GmInvqab{\mbf{q}_3}{\beta}{\beta_3}{*} \fdg{\mbf{q}_3 \beta_3}\right] \nonumber \\
\times & \left[\GmInvqab{\mbf{\tilde{q}_3}}{\beta}{\beta_4}{} f_{\mbf{\tilde{q}_3},\beta_4}+\GmInvqab{\mbf{{\tilde{q}}_3}}{\beta}{4+\beta_4}{} \fdg{\mbf{-\tilde{q}_3}, \beta_4}\right] \nonumber \\
& \textrm{where,~} \mbf{\tilde{q}_1} = \mbf{{q}_1 + Q} \textrm{~and~}, \mbf{\tilde{q}_3} = \mbf{{q}_3 - Q} \label{H_int_33_f_form}
\end{align}
}}}

\begin{figure*}[t]
\centering
\includegraphics[width=1.0\textwidth]{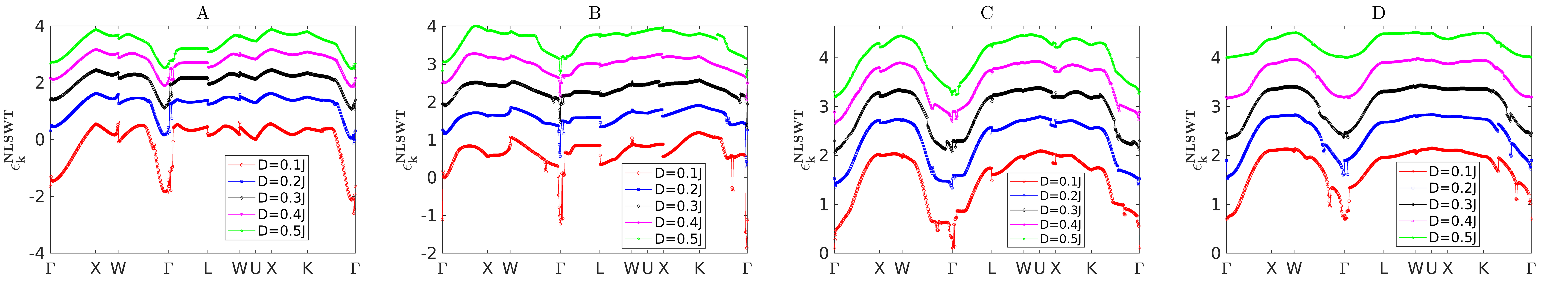}
\caption{{\small{{
{{\bf{(A, B, C, D)}}}: The renormalized spectrum for the four bands, from the lowest to the highest, evaluated using the self
energy at first order in nonlinear spin wave theory for $B\approx0 (=0.005J), S=1$ and
several strengths of the DMI interaction. For reference, the corresponding linear spin wave theory bands can be found in Fig. \ref{Fig_LSWT_spectra_panel}
}}}}
\label{Fig_ren_omega_all_bands}
\end{figure*}

Every quartic term in Eq. \ref{H_int_33_f_form} which has two creation and annihilation
operators can in principle contribute a non-vanishing term to the first order self energy.
One such term has the form 
$\fdg{\mbf{q}_1 \beta_1} f_{\mbf{q_1+Q},\beta_2} \fdg{\mbf{q}_3 \beta_3} f_{\mbf{q_3-Q},\beta_4}$. 
Writing this term in normal ordered form and applying Wick's theorem will result
in a contribution to the first order self energy from this term, which we call 
$ \mbf{\Sigma}^{1010, 33}_{\mu \nu}(q, \omega)$ (where $1\implies$ creation operator, $0\implies$ 
annihilation operator) of the form:

\begin{align}
&\mbf{\Sigma}^{1010, 33}_{\mu \nu}(\mbf{q}, \omega)= \frac{1}{N}{{\sumsbstck{\mbf{Q}}{\alpha, \beta, \beta_3}}} \jtab{3}{3}(\mbf{Q}) \times & \\
&\GmInvqab{\mbf{q}}{\alpha}{\mu}{*} 
\GmInvqab{\mbf{q+Q}}{\alpha}{\beta_3}{}
\GmInvqab{\mbf{q+Q}}{\beta}{\beta_3}{*}
\GmInvqab{\mbf{q}}{\beta}{\nu}{}  \nonumber \label{Sigma_33_1010}
\end{align}
There will clearly be several more contributions like this to the self energy from $H_{int}^{33}$.
The total irreducible self energy at first order, which we use for evaluation
of renormalized magnon bands in the rest of this work, is the sum of such contributions from all
the quartic terms in Eq. \ref{H_int_Eq}.
We note that at this order the self energy is frequency independent and Hermitian.
We emphasise that the notation we have used here is completely general and the above formalism 
is applicable for the analysis of renormalized magnon spectra of an ordered phase 
of a lattice with a basis provided the spin wave analysis is about a classical ground state
which has the translation invariance of the underlying Bravais lattice. We now present our results 
of this analysis for the AIAO phase of the pyrochlore lattice described by Eq. \ref{basic_H_test}.
\begin{figure*}[t]
\centering
\includegraphics[width=1.0\textwidth]{Figure_4_PRBPA}
\caption{{\small{{Spectral renormalisation with a magnetic field: (A, B, C). LSWT bands for a magnetic strength of $B=2.0J$ and
$D=0.3J$.
applied along the $[111]$ (A), $[100]$ (B) and $[110]$ (C), directions. (D, E, F). The renormalized spectra for the lowest band
for the same three directions for leading order spin wave interactions for $D\in[0.1J, 0.2J, 0.3J, 0.4J, 0.5J]$.
}}}}
\label{Fig_2_Bin80_data}
\end{figure*}

\section{Renormalised magnon spectra}
\label{ren_spectra}
For the evaluation of the renormalized spectra of the Hamiltonian we have used system sizes of $4 N^3$
sites with $N=10, 20$. For magnon spectra along a high symmetry path the evaluation of the leading order 
self energy involves a sum over momenta over the Brillouin zone. We evaluate that sum for the lattice 
sizes mentioned above and have verified that the difference in the final depicted spectra, a measure of 
finite size errors, is small. Throughout the text we show spectra along a path passing through high symmetry 
points in the Brillouin zone. Fig.\ref{Fig_BZ_Bdirs_GS}A depicts the Brillouin zone and the high symmetry 
points. \\ 

Before we discuss the renormalized spectra it is useful to have for reference LSWT spectra for different
$D$ and $B$ values. Fig. \ref{Fig_LSWT_spectra_panel} shows the magnonic band structure of the LSWT Hamiltonian 
for a vanishingly small magnetic field for $D=0.1J, 0.2J, 0.3J \textrm{~and~}0.4J$ (top row). 
We recall that an evaluation of the magnon bands of the all-in-all-out state in the absence of the DMI 
results in two sets of doubly degenerate bands, one flat band at zero energy and another dispersive 
\cite{Harris_Kallin_Berlinsky_PRB_1992}. The DMI results in a gap as can be seen in 
Fig. \ref{Fig_LSWT_spectra_panel} and also makes the lower bands dispersive.
We now investigate here what happens to these spectra under spin wave interactions.\\

At the level of first order in spin wave interactions the spectra are simply
renormalized but remain sharp. This follows from the frequency independence and
the hermiticity of the evaluated first order self energy matrices over the Brillouin zone.
At this leading order in magnon interactions we find significant renormalisation of 
the LSWT spectra. Fig \ref{Fig_ren_omega_all_bands} $(A, B, C, D)$ depict the renormalized bands starting
from the lowest to the highest for values of DMI in the range $D \in [0.1J, 0.5J]$. 
The general effect of the energy of the bands increasing as the value of DMI is increased remains 
true for the renormalized bands, however with substantial renormalisation. We also find that the
spectra in the vicinity of the $\Gamma$ point get renormalized to negative values for the lowest 
band indicating an instability of the AIAO phase to spin wave interactions. The data shown in 
Fig. \ref{Fig_ren_omega_all_bands} is depicted for $S=1$. We have also studied the Hamiltonian 
for $S=1/2$. The linear spin wave spectra are a simple linear function of the spin quantum number 
in the absence of a field. Hence all the magnon energies simply get renormalized to half their 
values in Fig. \ref{Fig_LSWT_spectra_panel} (top row). This results in the lowest band being nearer 
to the degenerate classical spin liquid manifold for $S=1/2$ than for $S=1$. Broadly, we find
that this results in the increased fragility of the AIAO phase for $S=1/2$ compared to $S=1$. 
The instability indicated by the negative values of the renormalized spectra are seen to be 
present for higher values of $D$ in the case of $S=1/2$ compared to $S=1$.
We note here that a tiny field of strength $0.005J$ was added to induce a small breaking of 
exact degeneracy in bands (which exists for $B=0$) and thence permit unambiguous band labels 
in the non-interacting Green's function. This also enables us to clearly identify bands with labels 
during the evaluation of the self energy, which involve a sum over the Brillouin zone wave vectors and bands. The data presented in Fig. \ref{Fig_ren_omega_all_bands}
is for a lattice of  $4N^3, N=20$ spins. \\ 

In Sec. \ref{model_and_formalism_section} we described the effect of magnetic fields on the classical ground state. We now present our results on the spectral renormalisation because of magnetic fields along the three
high symmetry directions. We have studied field strengths of $B=0.5J, 1.0J, 2.0J, 3.0J$ for
the all strengths of DMI we have analysed till now. These fields are considerably smaller than
the expected saturation fields for the DMI strengths studied (see Appendix \ref{canting_angle}).
For reference, a selection of non-interacting magnon spectra have been presented in 
Fig. \ref{Fig_LSWT_spectra_panel} (E, F, G, H). 

\begin{figure*}[t]
\centering
\includegraphics[width=1.0\textwidth]{Figure_5_PRBPA.png}
        \caption{{\small{{Spectral renormalisation of all bands as a function of the magnetic field: (A, B, C, D). LSWT bands (lowest to highest)
        for a magnetic field strengths $B \in [0.5J, 1.0J, 2.0J, 3.0J]$ and $D=0.3J$.
applied along the $[111]$ direction. (E, F, G, H). The renormalized spectra
for leading order spin wave interactions for the same parameters as in the figures in the top row.
}}}}
\label{Fig_3_Bvar_all_band_data}
\end{figure*}

Fig. \ref{Fig_2_Bin80_data}
depicts representative data showing the effects of a magnetic field. The top row depicts the LSWT spectra for $D=0.3J, \textrm{and}, B=2.0J$,
for $\mbf{B} \parallel [111], [100], [110]$ respectively. The bottom row depicts for various values of the strength of the DMI the 
renormalized spectra for the lowest band. 
A comparison between the plots in the bottom row and those of Fig. \ref{Fig_ren_omega_all_bands} clearly 
indicates that between the DMI and the field the former is the more important factor in determining the 
extent of the renormalisation for the parameter ranges that we have studied.
Both in Fig. \ref{Fig_ren_omega_all_bands} and in Fig. \ref{Fig_2_Bin80_data} we notice the 
appearance of several discontinuities and singularities in the renormalized spectrum. 
We find these features under one or more of several conditions. 
They often occur at high symmetry points and most prominently near the $\Gamma$ point. 
These discontinuities appear to have a regular trend in the vicinity of those points, in the sense that
the bands have a continuous behavior in either direction of the high symmetry point along the path. 
These indicate a discontinuous approach towards the high symmetry point from different directions in 
the Brillouin zone, within our evaluation method.  
A related kind of singular behavior happens to be at those wave vectors in the Brillouin zone where 
the LSWT spectra have point degeneracies.
In LSWT these are usually avoided crossings with very small gaps for the lattice sizes we study. 
Sometimes finite discontinuities like these can also
be found in regions of the Brillouin zone not necessarily close to any high symmetry point or small gaps in the LSWT spectra (see for instance the segment $K\Gamma$ in Fig. \ref{Fig_2_Bin80_data} F).
Considering the proximity of such singular behaviour in several cases to points in the Brillouin zone 
with small gaps, we believe many such jumps are the results of using the perturbative evaluation of the 
self energy beyond the limits of its applicability.
In the absence of such problems caused in a perturbative evaluation by small gaps in multiband systems
these kind of discontinuities have been reported in studies of magnon interactions in Bravais lattices, 
for instance in the triangular lattice \cite{SW_in_triangular_AFM_Chernyshev_Zhitomirsky_prb_2009}.
There they were found to be regions with an enhanced scattering rate for magnons. 
The fact that these features in the spectra are likely associated with an enhanced scattering rate 
is also indicated by the fact that an increase in the strength of DMI reduces the size of these effects 
as is clear in Fig \ref{Fig_2_Bin80_data} (D, E, F). An increase in DMI takes the single magnon spectrum 
further away from the degenerate classical manifold at $D=0$, thus reducing the 
available scattering space that is close to the band energy.
It is likely that a non-perturbative evaluation of the self energy and/or a higher order calculation 
will result in regular behaviour near such points.
However, since these jumps are at isolated points (in a three dimensional Brillouin zone) they do not 
result in substantial loss in accuracy for data that we present later for quantities over the full zone. 
We have not explored in this work possible methods to regularize the data near such points, such as 
introducing small terms in the Hamiltonian to increase the gap at such degeneracies. We note furthermore, 
that some of these point degeneracies are topological in nature and a proper treatment of interacting 
magnons around such points may not be possible using a perturbative framework.
\begin{figure*}[t]
\centering
\includegraphics[width=1.0\textwidth]{Figure_6_PRBPA.png}
\caption{{\small{{ The relative difference between the estimates of the ground state energy from linear and non-linear spin wave theory, with magnetic fields along three high symmetry directions and for various $D$ values.(a)-(c) The quantity plotted in all the three panels is $\frac{\mcl{E}_{GS}^{LSWT} - \mcl{E}_{GS}^{NLSWT}}{\vert \mcl{E}_{GS}^{LSWT}\vert}$, with magnetic field $B$ along the $[111]$, $[100]$ and $[110]$ directions respectively. }}}}
\label{Fig_4_E_gs_change}
\end{figure*}

As can be seen by a comparison of Fig. \ref{Fig_ren_omega_all_bands} and Fig. \ref{Fig_2_Bin80_data}, for the lowest band we notice that presence of a magnetic 
field has renormalized the bands further down for all the values of $D$ we have studied. It is relevant here to probe the effect on the renormalized spectra
by varying the field strength and also investigate the effects on the other bands. Fig. \ref{Fig_3_Bvar_all_band_data} presents both the non-interacting magnon spectra
as well as the renormalized spin wave spectra for several different values of the magnetic field and for all the four bands of the Hamiltonian. It is clear that for the lowest band
the stronger renormalisation towards lower values as the field strength is increased is generically true. Thus an increase in field strength for a given value of $D$ increases
the tendency towards destabilisation of the phase. We have shown in  Fig. \ref{Fig_3_Bvar_all_band_data} data for fields in the $[111]$ direction but the above statement
is true also for the other directions we have analyzed. The effect of the field however can depend on the band being considered. For instance, we see that the trend of decreasing renormalized spectra for an increasing 
field strength is reversed when we consider the highest band as can be seen in Fig. \ref{Fig_3_Bvar_all_band_data} H. The kind of non-analytic behaviors discussed for the lowest band earlier
are also present for the higher bands and as in the case of the lowest band the discontinuities decrease in magnitude as the strength of the DMI is increased.\\ 

\subsection{Ground state energy and renormalized magnon spectra}
\label{sec_gs_energy}

In the previous section we have been discussing the extent of spectral renormalisation because of magnon 
interactions along high symmetry paths in the Brillouin zone. We now analyze the effect of magnon interactions
on the spectra over the whole zone. The obvious way to extract this is to probe a quantity which depends over all
magnon frequencies. One such physical quantity with a simple dependence on magnon energies is the ground state energy. In the absence of magnon interactions we
know that the ground state energy of the non-interacting Hamiltonian is given by:
\beq
\mcl{E}_{GS}^{LSWT} = ~~\mcl{E}_0 +  {\sumlims{{\mbf{q}},\alpha}{}}^{\prime} \eta_{\mbf{q}}~  \epsilon_{-\mbf{q}\alpha}^{LSWT}
\label{E_GS_LSWT}
\eeq

where, $\dstl{\mcl{E}_0 \equiv N S(S+1) \sum_{\alpha \beta} \jtab{3}{3} (0) - N(S+\frac{1}{2}) \sum_{\alpha} B_{\alpha}^3}$ 
is a constant independent of the magnon frequencies. The $\prime$ over the summation indicates
that there is one term in the sum for each unique $(\mbf{q}, \mbf{-q})$ pair. $ \eta_{\mbf{q}}$ is $1$ for all wave vectors in the zone except for those points for which $\mbf{q}$ and $-\mbf{q}$ differ by a reciprocal lattice vector, in which case $ \eta_{\mbf{q}}=\frac{1}{2}$. Under spectral renormalisation, the renormalized spectra obtained from the poles of the Green's function are the single particle excitation energies of the interacting Hamiltonian. Replacing the non-interacting magnon energies in the above expression with the renormalized energies 
we can obtain using the change in the evaluated ground state energy an estimate of the effect of magnon interactions over the whole band. Thus,
\beq
\mcl{E}_{GS}^{NLSWT} = ~~\mcl{E}_0 +  {\sumlims{{\mbf{q}},\alpha}{}}^{\prime} \eta_{\mbf{q}}~  \epsilon_{-\mbf{q}\alpha}^{NLSWT}
\label{E_GS_NLSWT}
\eeq
can be considered to be the estimate of the ground state energy in the presence of the interactions.
Fig. \ref{Fig_4_E_gs_change} depicts the relative change in the ground state energy estimates because of magnon interactions.
Clearly, as expected from earlier plots a higher value of $D$ results in a smaller correction and hence a smaller overall effect
of the interaction terms. Interestingly, though for a given value of $D$ an increase in the strength of the magnetic field does increase
the system's tendency to be unstable by renormalising the lowest band to smaller values, overall for a bulk quantity like $\mcl{E}_{GS}$ the field seems to work against the spin wave interactions. 
For instance, in Fig. \ref{Fig_4_E_gs_change}, at any given $D$ increasing the value
of the field is taking the $\mcl{E}_{GS}^{NLSWT}$ value closer to $\mcl{E}_{GS}^{LSWT}$, the value
in the non-interacting limit. This essentially is a result of the fact that the field 
(unlike the DMI strength $D$) affects different bands differently and the bulk quantities 
being a sum over all bands feel the resulting cancellation effect.
\begin{figure*}[t]
\centering
\includegraphics[width=1.0\textwidth]{Figure_7_PRBPA.png}
\caption{{\small{{The difference between the bottom of the two magnon continuum at each wave vector and the band energy, minimized over the
whole Brillouin zone and all decay channels. A negative value indicates that the condition (Eq. \ref{two_MG_decay_condition}) for the magnon
decay is satisfied, for the band considered, somewhere in the Brillouin zone. A, B, C, D are the data
for the first (lowest), second, third and fourth band respectively. Data presented is evaluated for a lattice size of $4N^3 (N=20)$ spins.}}}}
\label{Fig_5_two_MG_min_delta}
\end{figure*}
\section{Two magnon continuum and magnon decay}
In the previous section we have analyzed the poles of the single particle Green's function by
evaluating the leading order term of the self energy. 
At this order, as mentioned in the previous section, the poles
are real and hence magnon interactions result only in the renormalisation of the LSWT spectra while
keeping the excitations sharp. The lowest order term which can result in poles with an imaginary part
is at second order of terms involving the three bosons, for e.g. $\jtab{3}{+}$ etc. in Eq. \ref{H_int_Eq}. 

In principle, such a second order term can result in magnon decay via scattering into two magnons 
\cite{magnon_decay_RMP_2013}. 
A detailed analysis of the pole structure and the associated imaginary parts requires the
evaluation of the second order self energy and 
a numerical search for the singularities of the Green's function using the evaluated self energy
till the second order. However, it is possible to obtain preliminary information about the 
approximate energy regions
where magnon decay is allowed using the second order self energy at the LSWT energies.
In that approximation in order for magnon decay to be possible 
the LSWT spectra have to satisfy certain kinematic conditions 
\cite{magnon_decay_RMP_2013, Starykh_Chubukov_Abanov_PRB_2006, Chernyshev_Maksimov_Honeycomb_PRB_2016}.
For instance, for the decay of the $\alpha$-th band at the wave vector $k$ into two magnons in the $\beta$-th and $\gamma$-th bands,
requires: 
\beq
\epsilon_{\mbf{k}, \alpha} = \epsilon_{\mbf{q}, \beta} + \epsilon_{\mbf{k-q}, \gamma}
\label{two_MG_decay_condition}
\eeq
for some $\mbf{q}$ in the Brillouin zone. Thus, there should be an intersection of the band of two magnon excitations
corresponding to a given momentum
with the single magnon band at the same wave vector for such a decay to be possible. A specific triplet $(\alpha, \beta, \gamma)$ 
in Eq. \ref{two_MG_decay_condition} is a particular decay channel and a given band $\alpha$ can decay via multiple channels. 
We have investigated this possibility of decay for the system we are studying and the results are displayed in Fig. \ref{Fig_5_two_MG_min_delta}.\\

The quantity $\Delta^{two-magnon}_{min}$ which has been plotted is defined in the following manner:
\beq
\Delta^{two-magnon}_{min} (\alpha) = \min_{\mbf{k,q, \beta, \gamma}}{\left[(\epsilon_{\mbf{q}, \beta} + \epsilon_{\mbf{k-q}, \gamma}) -\epsilon_{\mbf{k}, \alpha} \right]}
\label{delta_min_def}
\eeq
where $\mbf{k, q}$ are both vectors in the first Brillouin zone.
Thus $ \Delta^{two-magnon}_{min} (\alpha) $  quantity is related to the gap between the bottom of the two magnon continuum and the energy of the $\alpha$-th band
at wave vectors in the Brillouin zone. The minimization over $(k,\beta, \gamma)$ indicates, should the minimum be a negative number, that the two magnon continuum and
the band intersect somewhere in the Brillouin zone. Thus Fig. \ref{Fig_5_two_MG_min_delta} depicts a one-parameter diagnostic, for the ranges of $B$ and $D$ we 
have studied, to indicate if a magnon decay is possible.\\
The data clearly demonstrates that the lowest band, if it is not destabilized by the leading order interaction term, does not satisfy the kinematic condition  
for decay for the range of fields and DMI interactions that
we have studied. Furthermore, we see clearly that the tendency to decay increases for any given strength of DMI with the increase
of the strength of the magnetic field and decreases as the strength of the DMI is increased. We note here that the condition Eq. \ref{two_MG_decay_condition}
is a necessary, but not a sufficient condition for magnon decay through this mechanism. A non-vanishing self energy matrix element will also require
the relevant matrix element dependent on $\jtab{3}{+}$ etc. to be non-zero. Thus the results of Fig. \ref{Fig_5_two_MG_min_delta} should be understood
as being indicative of the parameters where the bands are protected against decay and possible parameter regions where decay might occur, provided other conditions
are satisfied. 

\section{Summary and discussion}
\label{discussion_section}
The all-in-all-out phase of the pyrochlore lattice is a very suitable
platform for probing the effect of spin wave interactions. 
It is a non-collinear ground state, which can be close in parameter
and energy space to a macroscopically degenerate ground state manifold of a geometrically frustrated system.
In this preliminary analysis of the effect of spin wave interactions in this phase we have shown that the
stability of the phase can be jeopardized below some strength of the DMI. Also, the application of a magnetic
field generically increases a tendency towards destabilisation of the phase. The nature of that phase itself
cannot be gauged from our analysis but for our parameter values it is far from the field dominated ordered phases.

Our results in this work are restricted to the leading order calculations of self energy. 
The next order effects have in several cases been accounted for using an effective modification 
to the three boson terms of the interaction Hamiltonian and the associated canting angle renormalisation 
\cite{Zhitomirsky_and_Nikuni_prb_1998, Zhitomirsky_and_Chernyshev_sq_lat_PRL_1999}. 
Such higher order effects, though not negligible, are smaller than the leading order effects for fields
far away from saturation \cite{Zhitomirsky_and_Chernyshev_sq_lat_PRL_1999}. In our study, which is
on a three dimensional system with a multiple sublattice structure, we have restricted our analysis
to the leading order, considering the higher dimension and the values of the field being far from staturation.
On general grounds, considering the sub-dominant effect at low fields in two dimensions, we expect 
higher order effects to be weak compared to the leading order effects for those of our parameter regimes
where order is stable up to spectral renormalisation. Our objective in this work has been to demonstrate 
the trend of spectral renormalisation in a broad range of parameters for a general class of 
Hamiltonians on the pyrochlore lattice. 
While the general trends with change in DMI or the 
fields can be expected to be the same, the precise extent of band renormalisation
for a given value of DMI and field can differ because of higher order effects.

In this study we have also not addressed the issue of the fate of the topological nature of the band 
structures with the AIAO phase, which might require non-perturbative techniques close to topological degeneracies as mentioned in the text. 
However, the significant spectral renormalisation we report especially along high symmetry directions in the
Brillouin zone means that the estimation of the phase boundaries for transitions to such phases might be quite different
from those gauged from linear spin wave theory. 
Finally, our evaluations of the possible parameter
regions to probe magnon decay can be extended to give detailed accounts of the 
channel resolved decay surfaces or the quantitative effect on scattering cross-sections. 
Having established the necessity to take spin wave interactions into account, these more extensive 
analyses are among the logical next steps for this phase of the pyrochlore lattice.

\revappendix
\counterwithin{figure}{section}

\section{Variation of canting angle with the field}
\label{canting_angle}

\begin{figure*}[t]
\centering
\includegraphics[width=1.0\textwidth]{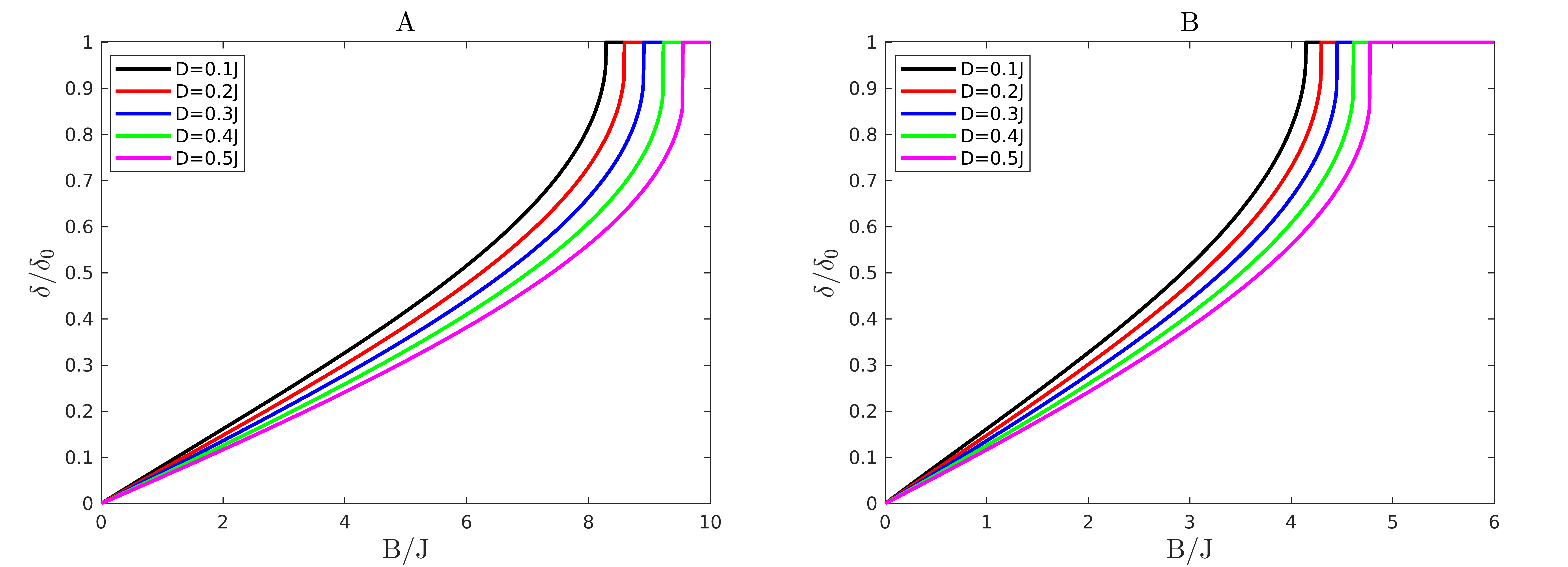}
\caption{{\small{{
Variation of the canting (tilt) $\delta$ towards the field direction relative to the moment orientation
in the AIAO state, when the field is along the $[111]$ direction. $\delta$ is effectively the angle between the
green and red arrows in Fig. \ref{Fig_BZ_Bdirs_GS} for those spins which are not aligned with the field direction. $\delta_0$ is the angle between the AIAO direction and the field direction. $\delta / \delta_0 =1$ implies the state fully polarized along the field direction. A. $S=1$ B. $S=1/2$.
}}}}
\label{Fig_canting_angle}
\end{figure*}
In the main text we discussed the effect of a finite magnetic field on the classical AIAO state
which is the lowest energy state in the absence of a field. In Fig. \ref{Fig_BZ_Bdirs_GS}C  
we depicted the nature of the classical ground state for a field along the $[111]$ direction.
In this case, the three spins of the tetrahedron not aligned with the field tilt towards the
field by the same angle $\delta$ and the canting happens for each spin in the plane 
containing the field direction and the initial orientation of the moment. 
Using these as the defining features of the ground state we can reduce
the problem of finding the classical ground state to minimising the following
classical energy function for a given value of $J, D \textrm{~and~} B$.
\beqa
E^{classical}_{per-site} = \frac{3JS^2}{8} - \frac{3\sqrt{2}DS^2}{8} - \frac{BS}{4} \nonumber \\
+ \sin(\delta) \left [\sqrt{2}JS^2 - \frac{DS^2}{{2}} - \frac{BS}{\sqrt{2}} \right] \nonumber \\
+ \cos(\delta) \left[ \frac{BS}{4}  - \sqrt{2} D S^2- \frac{J S^2}{2} \right] \nonumber \\
+ \sin(2 \delta) \left[\frac{D S^2}{4}-\frac{J S^2}{\sqrt{2}}\right] \nonumber \\
+ \cos(2 \delta) \left[-\frac{5 \sqrt{2} D S^2}{8}-\frac{7 J S^2}{8}\right] \label{Eq_AIAO_cl_gs} 
\eeqa

The tilt angle corresponding to the classical ground state is the $\delta$ corresponding to the minimum of the $E^{classical}_{per-site}$. It is possible to obtain estimates of the saturation field for this class of states
by evaluation of the minima for a range of magnetic fields. The results are shown in Fig. \ref{Fig_canting_angle} 
both for $S=1 \textrm{~and~} S=1/2$. As expected the canting decreases with the increase in the strength
of the DMI and for the figures depicted in the main text the fields are well below the saturation fields.
The results of the minimisation of the above function do coincide with the numerical minimisation results
(produced without using any constraints on the spin orientations) used in the main text.
We point out here that the above estimates of the saturation fields rely on the specific nature of the 
magnetic state used, namely coplanarity of the initial and final direction of the moment and the magnetic field
direction. We do not know of any proof that the class of minimising functions will satisfy this property
for the entire range of fields depicted in Fig. \ref{Fig_canting_angle}. However, as already mentioned, we
rely on unconstrained minimisation and for our field values of interest the property holds. 
An expression like Eq.\ref {Eq_AIAO_cl_gs} can in principle be written for the $[100]$ direction as well using 
two different canting angles but as it should be clear to the reader it is likely to be considerably more 
involved and we do not provide explicit expressions for the same.
\acknowledgments
V.R.C acknowledges funding from the Department of Atomic Energy, India under the project number RIN 4001-SPS.


\bibliography{references_database_paper}

\begin{thebibliography}{45}%
\makeatletter
\providecommand \@ifxundefined [1]{%
 \@ifx{#1\undefined}
}%
\providecommand \@ifnum [1]{%
 \ifnum #1\expandafter \@firstoftwo
 \else \expandafter \@secondoftwo
 \fi
}%
\providecommand \@ifx [1]{%
 \ifx #1\expandafter \@firstoftwo
 \else \expandafter \@secondoftwo
 \fi
}%
\providecommand \natexlab [1]{#1}%
\providecommand \enquote  [1]{``#1''}%
\providecommand \bibnamefont  [1]{#1}%
\providecommand \bibfnamefont [1]{#1}%
\providecommand \citenamefont [1]{#1}%
\providecommand \href@noop [0]{\@secondoftwo}%
\providecommand \href [0]{\begingroup \@sanitize@url \@href}%
\providecommand \@href[1]{\@@startlink{#1}\@@href}%
\providecommand \@@href[1]{\endgroup#1\@@endlink}%
\providecommand \@sanitize@url [0]{\catcode `\\12\catcode `\$12\catcode
  `\&12\catcode `\#12\catcode `\^12\catcode `\_12\catcode `\%12\relax}%
\providecommand \@@startlink[1]{}%
\providecommand \@@endlink[0]{}%
\providecommand \url  [0]{\begingroup\@sanitize@url \@url }%
\providecommand \@url [1]{\endgroup\@href {#1}{\urlprefix }}%
\providecommand \urlprefix  [0]{URL }%
\providecommand \Eprint [0]{\href }%
\providecommand \doibase [0]{https://doi.org/}%
\providecommand \selectlanguage [0]{\@gobble}%
\providecommand \bibinfo  [0]{\@secondoftwo}%
\providecommand \bibfield  [0]{\@secondoftwo}%
\providecommand \translation [1]{[#1]}%
\providecommand \BibitemOpen [0]{}%
\providecommand \bibitemStop [0]{}%
\providecommand \bibitemNoStop [0]{.\EOS\space}%
\providecommand \EOS [0]{\spacefactor3000\relax}%
\providecommand \BibitemShut  [1]{\csname bibitem#1\endcsname}%
\let\auto@bib@innerbib\@empty
\bibitem [{\citenamefont {{Bloch}}(1930)}]{Bloch_SW_1930}%
  \BibitemOpen
  \bibfield  {author} {\bibinfo {author} {\bibfnamefont {F.}~\bibnamefont
  {{Bloch}}},\ }\bibfield  {title} {\bibinfo {title} {{Zur Theorie des
  Ferromagnetismus}},\ }\href {https://doi.org/10.1007/BF01339661} {\bibfield
  {journal} {\bibinfo  {journal} {Zeitschrift fur Physik}\ }\textbf {\bibinfo
  {volume} {61}},\ \bibinfo {pages} {206} (\bibinfo {year} {1930})}\BibitemShut
  {NoStop}%
\bibitem [{\citenamefont {Anderson}(1952)}]{Anderson_SW_1952}%
  \BibitemOpen
  \bibfield  {author} {\bibinfo {author} {\bibfnamefont {P.~W.}\ \bibnamefont
  {Anderson}},\ }\bibfield  {title} {\bibinfo {title} {An approximate quantum
  theory of the antiferromagnetic ground state},\ }\href
  {https://doi.org/10.1103/PhysRev.86.694} {\bibfield  {journal} {\bibinfo
  {journal} {Phys. Rev.}\ }\textbf {\bibinfo {volume} {86}},\ \bibinfo {pages}
  {694} (\bibinfo {year} {1952})}\BibitemShut {NoStop}%
\bibitem [{\citenamefont {Kubo}(1952)}]{Kubo_1952_SW}%
  \BibitemOpen
  \bibfield  {author} {\bibinfo {author} {\bibfnamefont {R.}~\bibnamefont
  {Kubo}},\ }\bibfield  {title} {\bibinfo {title} {The spin-wave theory of
  antiferromagnetics},\ }\href {https://doi.org/10.1103/PhysRev.87.568}
  {\bibfield  {journal} {\bibinfo  {journal} {Phys. Rev.}\ }\textbf {\bibinfo
  {volume} {87}},\ \bibinfo {pages} {568} (\bibinfo {year} {1952})}\BibitemShut
  {NoStop}%
\bibitem [{\citenamefont
  {Dyson}(1956{\natexlab{a}})}]{Dyson_SW_FM_interactions_1}%
  \BibitemOpen
  \bibfield  {author} {\bibinfo {author} {\bibfnamefont {F.~J.}\ \bibnamefont
  {Dyson}},\ }\bibfield  {title} {\bibinfo {title} {General theory of spin-wave
  interactions},\ }\href {https://doi.org/10.1103/PhysRev.102.1217} {\bibfield
  {journal} {\bibinfo  {journal} {Phys. Rev.}\ }\textbf {\bibinfo {volume}
  {102}},\ \bibinfo {pages} {1217} (\bibinfo {year}
  {1956}{\natexlab{a}})}\BibitemShut {NoStop}%
\bibitem [{\citenamefont
  {Dyson}(1956{\natexlab{b}})}]{Dyson_SW_FM_interactions_2}%
  \BibitemOpen
  \bibfield  {author} {\bibinfo {author} {\bibfnamefont {F.~J.}\ \bibnamefont
  {Dyson}},\ }\bibfield  {title} {\bibinfo {title} {Thermodynamic behavior of
  an ideal ferromagnet},\ }\href {https://doi.org/10.1103/PhysRev.102.1230}
  {\bibfield  {journal} {\bibinfo  {journal} {Phys. Rev.}\ }\textbf {\bibinfo
  {volume} {102}},\ \bibinfo {pages} {1230} (\bibinfo {year}
  {1956}{\natexlab{b}})}\BibitemShut {NoStop}%
\bibitem [{\citenamefont {Oguchi}(1960)}]{Oguchi_SW_interactions_1960}%
  \BibitemOpen
  \bibfield  {author} {\bibinfo {author} {\bibfnamefont {T.}~\bibnamefont
  {Oguchi}},\ }\bibfield  {title} {\bibinfo {title} {Theory of spin-wave
  interactions in ferro- and antiferromagnetism},\ }\href
  {https://doi.org/10.1103/PhysRev.117.117} {\bibfield  {journal} {\bibinfo
  {journal} {Phys. Rev.}\ }\textbf {\bibinfo {volume} {117}},\ \bibinfo {pages}
  {117} (\bibinfo {year} {1960})}\BibitemShut {NoStop}%
\bibitem [{\citenamefont {Chernyshev}\ and\ \citenamefont
  {Zhitomirsky}(2009)}]{SW_in_triangular_AFM_Chernyshev_Zhitomirsky_prb_2009}%
  \BibitemOpen
  \bibfield  {author} {\bibinfo {author} {\bibfnamefont {A.~L.}\ \bibnamefont
  {Chernyshev}}\ and\ \bibinfo {author} {\bibfnamefont {M.~E.}\ \bibnamefont
  {Zhitomirsky}},\ }\bibfield  {title} {\bibinfo {title} {Spin waves in a
  triangular lattice antiferromagnet: Decays, spectrum renormalization, and
  singularities},\ }\href {https://doi.org/10.1103/PhysRevB.79.144416}
  {\bibfield  {journal} {\bibinfo  {journal} {Phys. Rev. B}\ }\textbf {\bibinfo
  {volume} {79}},\ \bibinfo {pages} {144416} (\bibinfo {year}
  {2009})}\BibitemShut {NoStop}%
\bibitem [{\citenamefont {Zhitomirsky}\ and\ \citenamefont
  {Chernyshev}(2013)}]{magnon_decay_RMP_2013}%
  \BibitemOpen
  \bibfield  {author} {\bibinfo {author} {\bibfnamefont {M.~E.}\ \bibnamefont
  {Zhitomirsky}}\ and\ \bibinfo {author} {\bibfnamefont {A.~L.}\ \bibnamefont
  {Chernyshev}},\ }\bibfield  {title} {\bibinfo {title} {Colloquium:
  Spontaneous magnon decays},\ }\href
  {https://doi.org/10.1103/RevModPhys.85.219} {\bibfield  {journal} {\bibinfo
  {journal} {Rev. Mod. Phys.}\ }\textbf {\bibinfo {volume} {85}},\ \bibinfo
  {pages} {219} (\bibinfo {year} {2013})}\BibitemShut {NoStop}%
\bibitem [{\citenamefont {Gardner}\ \emph {et~al.}(2010)\citenamefont
  {Gardner}, \citenamefont {Gingras},\ and\ \citenamefont
  {Greedan}}]{gardner_gingras_greedan_rmp}%
  \BibitemOpen
  \bibfield  {author} {\bibinfo {author} {\bibfnamefont {J.~S.}\ \bibnamefont
  {Gardner}}, \bibinfo {author} {\bibfnamefont {M.~J.~P.}\ \bibnamefont
  {Gingras}},\ and\ \bibinfo {author} {\bibfnamefont {J.~E.}\ \bibnamefont
  {Greedan}},\ }\bibfield  {title} {\bibinfo {title} {Magnetic pyrochlore
  oxides},\ }\href {https://doi.org/10.1103/RevModPhys.82.53} {\bibfield
  {journal} {\bibinfo  {journal} {Rev. Mod. Phys.}\ }\textbf {\bibinfo {volume}
  {82}},\ \bibinfo {pages} {53} (\bibinfo {year} {2010})}\BibitemShut {NoStop}%
\bibitem [{\citenamefont {Onose}\ \emph {et~al.}(2010)\citenamefont {Onose},
  \citenamefont {Ideue}, \citenamefont {Katsura}, \citenamefont {Shiomi},
  \citenamefont {Nagaosa},\ and\ \citenamefont
  {Tokura}}]{onose_et_al_science_2010}%
  \BibitemOpen
  \bibfield  {author} {\bibinfo {author} {\bibfnamefont {Y.}~\bibnamefont
  {Onose}}, \bibinfo {author} {\bibfnamefont {T.}~\bibnamefont {Ideue}},
  \bibinfo {author} {\bibfnamefont {H.}~\bibnamefont {Katsura}}, \bibinfo
  {author} {\bibfnamefont {Y.}~\bibnamefont {Shiomi}}, \bibinfo {author}
  {\bibfnamefont {N.}~\bibnamefont {Nagaosa}},\ and\ \bibinfo {author}
  {\bibfnamefont {Y.}~\bibnamefont {Tokura}},\ }\bibfield  {title} {\bibinfo
  {title} {Observation of the magnon hall effect},\ }\href
  {https://doi.org/10.1126/science.1188260} {\bibfield  {journal} {\bibinfo
  {journal} {Science}\ }\textbf {\bibinfo {volume} {329}},\ \bibinfo {pages}
  {297} (\bibinfo {year} {2010})}\BibitemShut {NoStop}%
\bibitem [{\citenamefont {McClarty}(2022)}]{McClarty_Top_Mag_review}%
  \BibitemOpen
  \bibfield  {author} {\bibinfo {author} {\bibfnamefont {P.~A.}\ \bibnamefont
  {McClarty}},\ }\bibfield  {title} {\bibinfo {title} {Topological magnons: A
  review},\ }\href {https://doi.org/10.1146/annurev-conmatphys-031620-104715}
  {\bibfield  {journal} {\bibinfo  {journal} {Annual Review of Condensed Matter
  Physics}\ }\textbf {\bibinfo {volume} {13}},\ \bibinfo {pages} {171}
  (\bibinfo {year} {2022})},\ \Eprint
  {https://arxiv.org/abs/https://doi.org/10.1146/annurev-conmatphys-031620-104715}
  {https://doi.org/10.1146/annurev-conmatphys-031620-104715} \BibitemShut
  {NoStop}%
\bibitem [{\citenamefont {Malki}\ and\ \citenamefont
  {Uhrig}(2020)}]{Malki_top_mag_review_2020}%
  \BibitemOpen
  \bibfield  {author} {\bibinfo {author} {\bibfnamefont {M.}~\bibnamefont
  {Malki}}\ and\ \bibinfo {author} {\bibfnamefont {G.~S.}\ \bibnamefont
  {Uhrig}},\ }\bibfield  {title} {\bibinfo {title} {Topological magnetic
  excitations},\ }\href {https://doi.org/10.1209/0295-5075/132/20003}
  {\bibfield  {journal} {\bibinfo  {journal} {Europhysics Letters}\ }\textbf
  {\bibinfo {volume} {132}},\ \bibinfo {pages} {20003} (\bibinfo {year}
  {2020})}\BibitemShut {NoStop}%
\bibitem [{\citenamefont {Kondo}\ \emph {et~al.}(2020)\citenamefont {Kondo},
  \citenamefont {Akagi},\ and\ \citenamefont
  {Katsura}}]{kondo_akagi_katsura_2020_ptep}%
  \BibitemOpen
  \bibfield  {author} {\bibinfo {author} {\bibfnamefont {H.}~\bibnamefont
  {Kondo}}, \bibinfo {author} {\bibfnamefont {Y.}~\bibnamefont {Akagi}},\ and\
  \bibinfo {author} {\bibfnamefont {H.}~\bibnamefont {Katsura}},\ }\bibfield
  {title} {\bibinfo {title} {{Non-Hermiticity and topological invariants of
  magnon Bogoliubov–de Gennes systems}},\ }\bibfield  {journal} {\bibinfo
  {journal} {Progress of Theoretical and Experimental Physics}\ }\textbf
  {\bibinfo {volume} {2020}},\ \href {https://doi.org/10.1093/ptep/ptaa151}
  {10.1093/ptep/ptaa151} (\bibinfo {year} {2020})\BibitemShut {NoStop}%
\bibitem [{\citenamefont {Chernyshev}\ and\ \citenamefont
  {Maksimov}(2016)}]{dampped_TM_in_Kagome_Chernyshev_Maximov_prl_2016}%
  \BibitemOpen
  \bibfield  {author} {\bibinfo {author} {\bibfnamefont {A.~L.}\ \bibnamefont
  {Chernyshev}}\ and\ \bibinfo {author} {\bibfnamefont {P.~A.}\ \bibnamefont
  {Maksimov}},\ }\bibfield  {title} {\bibinfo {title} {Damped topological
  magnons in the kagome-lattice ferromagnets},\ }\href
  {https://doi.org/10.1103/PhysRevLett.117.187203} {\bibfield  {journal}
  {\bibinfo  {journal} {Phys. Rev. Lett.}\ }\textbf {\bibinfo {volume} {117}},\
  \bibinfo {pages} {187203} (\bibinfo {year} {2016})}\BibitemShut {NoStop}%
\bibitem [{\citenamefont {Pershoguba}\ \emph {et~al.}(2018)\citenamefont
  {Pershoguba}, \citenamefont {Banerjee}, \citenamefont {Lashley},
  \citenamefont {Park}, \citenamefont {\AA{}gren}, \citenamefont {Aeppli},\
  and\ \citenamefont {Balatsky}}]{Pershoguba_et_PRX_2018}%
  \BibitemOpen
  \bibfield  {author} {\bibinfo {author} {\bibfnamefont {S.~S.}\ \bibnamefont
  {Pershoguba}}, \bibinfo {author} {\bibfnamefont {S.}~\bibnamefont
  {Banerjee}}, \bibinfo {author} {\bibfnamefont {J.~C.}\ \bibnamefont
  {Lashley}}, \bibinfo {author} {\bibfnamefont {J.}~\bibnamefont {Park}},
  \bibinfo {author} {\bibfnamefont {H.}~\bibnamefont {\AA{}gren}}, \bibinfo
  {author} {\bibfnamefont {G.}~\bibnamefont {Aeppli}},\ and\ \bibinfo {author}
  {\bibfnamefont {A.~V.}\ \bibnamefont {Balatsky}},\ }\bibfield  {title}
  {\bibinfo {title} {Dirac magnons in honeycomb ferromagnets},\ }\href
  {https://doi.org/10.1103/PhysRevX.8.011010} {\bibfield  {journal} {\bibinfo
  {journal} {Phys. Rev. X}\ }\textbf {\bibinfo {volume} {8}},\ \bibinfo {pages}
  {011010} (\bibinfo {year} {2018})}\BibitemShut {NoStop}%
\bibitem [{\citenamefont {Mook}\ \emph {et~al.}(2021)\citenamefont {Mook},
  \citenamefont {Plekhanov}, \citenamefont {Klinovaja},\ and\ \citenamefont
  {Loss}}]{mook_et_al_interaction_stabilised_TMI_2022}%
  \BibitemOpen
  \bibfield  {author} {\bibinfo {author} {\bibfnamefont {A.}~\bibnamefont
  {Mook}}, \bibinfo {author} {\bibfnamefont {K.}~\bibnamefont {Plekhanov}},
  \bibinfo {author} {\bibfnamefont {J.}~\bibnamefont {Klinovaja}},\ and\
  \bibinfo {author} {\bibfnamefont {D.}~\bibnamefont {Loss}},\ }\bibfield
  {title} {\bibinfo {title} {Interaction-stabilized topological magnon
  insulator in ferromagnets},\ }\href
  {https://doi.org/10.1103/PhysRevX.11.021061} {\bibfield  {journal} {\bibinfo
  {journal} {Phys. Rev. X}\ }\textbf {\bibinfo {volume} {11}},\ \bibinfo
  {pages} {021061} (\bibinfo {year} {2021})}\BibitemShut {NoStop}%
\bibitem [{\citenamefont {Liu}\ \emph {et~al.}(2023)\citenamefont {Liu},
  \citenamefont {Zhai}, \citenamefont {Yan}, \citenamefont {Wang},\ and\
  \citenamefont {Wan}}]{Liu_et_al_SW_interactions_MnBi2Te4_prb_2023}%
  \BibitemOpen
  \bibfield  {author} {\bibinfo {author} {\bibfnamefont {Y.}~\bibnamefont
  {Liu}}, \bibinfo {author} {\bibfnamefont {L.}~\bibnamefont {Zhai}}, \bibinfo
  {author} {\bibfnamefont {S.}~\bibnamefont {Yan}}, \bibinfo {author}
  {\bibfnamefont {D.}~\bibnamefont {Wang}},\ and\ \bibinfo {author}
  {\bibfnamefont {X.}~\bibnamefont {Wan}},\ }\bibfield  {title} {\bibinfo
  {title} {Magnon-magnon interaction in monolayer
  ${\mathrm{mnbi}}_{2}{\mathrm{te}}_{4}$},\ }\href
  {https://doi.org/10.1103/PhysRevB.108.174425} {\bibfield  {journal} {\bibinfo
   {journal} {Phys. Rev. B}\ }\textbf {\bibinfo {volume} {108}},\ \bibinfo
  {pages} {174425} (\bibinfo {year} {2023})}\BibitemShut {NoStop}%
\bibitem [{\citenamefont {Chen}\ \emph {et~al.}(2023)\citenamefont {Chen},
  \citenamefont {Huang},\ and\ \citenamefont
  {Fu}}]{honeycomb_damping_chen_etal_prb_2023}%
  \BibitemOpen
  \bibfield  {author} {\bibinfo {author} {\bibfnamefont {Q.-H.}\ \bibnamefont
  {Chen}}, \bibinfo {author} {\bibfnamefont {F.-J.}\ \bibnamefont {Huang}},\
  and\ \bibinfo {author} {\bibfnamefont {Y.-P.}\ \bibnamefont {Fu}},\
  }\bibfield  {title} {\bibinfo {title} {Damped topological magnons in
  honeycomb antiferromagnets},\ }\href
  {https://doi.org/10.1103/PhysRevB.108.024409} {\bibfield  {journal} {\bibinfo
   {journal} {Phys. Rev. B}\ }\textbf {\bibinfo {volume} {108}},\ \bibinfo
  {pages} {024409} (\bibinfo {year} {2023})}\BibitemShut {NoStop}%
\bibitem [{\citenamefont {Habel}\ \emph {et~al.}(2024)\citenamefont {Habel},
  \citenamefont {Mook}, \citenamefont {Willsher},\ and\ \citenamefont
  {Knolle}}]{Habel_et_al_PRB_2023}%
  \BibitemOpen
  \bibfield  {author} {\bibinfo {author} {\bibfnamefont {J.}~\bibnamefont
  {Habel}}, \bibinfo {author} {\bibfnamefont {A.}~\bibnamefont {Mook}},
  \bibinfo {author} {\bibfnamefont {J.}~\bibnamefont {Willsher}},\ and\
  \bibinfo {author} {\bibfnamefont {J.}~\bibnamefont {Knolle}},\ }\bibfield
  {title} {\bibinfo {title} {Breakdown of chiral edge modes in topological
  magnon insulators},\ }\href {https://doi.org/10.1103/PhysRevB.109.024441}
  {\bibfield  {journal} {\bibinfo  {journal} {Phys. Rev. B}\ }\textbf {\bibinfo
  {volume} {109}},\ \bibinfo {pages} {024441} (\bibinfo {year}
  {2024})}\BibitemShut {NoStop}%
\bibitem [{\citenamefont {Sun}\ \emph {et~al.}(2023)\citenamefont {Sun},
  \citenamefont {Bhowmick}, \citenamefont {Yang},\ and\ \citenamefont
  {Sengupta}}]{Sun_Bhowmick_Yang_Sengupta_PRB_2023}%
  \BibitemOpen
  \bibfield  {author} {\bibinfo {author} {\bibfnamefont {H.}~\bibnamefont
  {Sun}}, \bibinfo {author} {\bibfnamefont {D.}~\bibnamefont {Bhowmick}},
  \bibinfo {author} {\bibfnamefont {B.}~\bibnamefont {Yang}},\ and\ \bibinfo
  {author} {\bibfnamefont {P.}~\bibnamefont {Sengupta}},\ }\bibfield  {title}
  {\bibinfo {title} {Interacting topological dirac magnons},\ }\href
  {https://doi.org/10.1103/PhysRevB.107.134426} {\bibfield  {journal} {\bibinfo
   {journal} {Phys. Rev. B}\ }\textbf {\bibinfo {volume} {107}},\ \bibinfo
  {pages} {134426} (\bibinfo {year} {2023})}\BibitemShut {NoStop}%
\bibitem [{\citenamefont {Sourounis}\ and\ \citenamefont
  {Manchon}(2024)}]{sourounis2024PRB}%
  \BibitemOpen
  \bibfield  {author} {\bibinfo {author} {\bibfnamefont {K.}~\bibnamefont
  {Sourounis}}\ and\ \bibinfo {author} {\bibfnamefont {A.}~\bibnamefont
  {Manchon}},\ }\bibfield  {title} {\bibinfo {title} {Impact of magnon
  interactions on transport in honeycomb antiferromagnets},\ }\href
  {https://doi.org/10.1103/PhysRevB.110.054429} {\bibfield  {journal} {\bibinfo
   {journal} {Phys. Rev. B}\ }\textbf {\bibinfo {volume} {110}},\ \bibinfo
  {pages} {054429} (\bibinfo {year} {2024})}\BibitemShut {NoStop}%
\bibitem [{\citenamefont {Li}\ \emph {et~al.}(2023)\citenamefont {Li},
  \citenamefont {Luo},\ and\ \citenamefont
  {Chang}}]{Li_Luo_Chang_T_induced_Chern_insulator_PRB_2023}%
  \BibitemOpen
  \bibfield  {author} {\bibinfo {author} {\bibfnamefont {Y.-M.}\ \bibnamefont
  {Li}}, \bibinfo {author} {\bibfnamefont {X.-W.}\ \bibnamefont {Luo}},\ and\
  \bibinfo {author} {\bibfnamefont {K.}~\bibnamefont {Chang}},\ }\bibfield
  {title} {\bibinfo {title} {Temperature-induced magnonic chern insulator in
  collinear antiferromagnets},\ }\href
  {https://doi.org/10.1103/PhysRevB.107.214417} {\bibfield  {journal} {\bibinfo
   {journal} {Phys. Rev. B}\ }\textbf {\bibinfo {volume} {107}},\ \bibinfo
  {pages} {214417} (\bibinfo {year} {2023})}\BibitemShut {NoStop}%
\bibitem [{\citenamefont {Rau}\ \emph {et~al.}(2019)\citenamefont {Rau},
  \citenamefont {Moessner},\ and\ \citenamefont
  {McClarty}}]{Rau_Moessner_McClarty_NLSWT_PRB_2019}%
  \BibitemOpen
  \bibfield  {author} {\bibinfo {author} {\bibfnamefont {J.~G.}\ \bibnamefont
  {Rau}}, \bibinfo {author} {\bibfnamefont {R.}~\bibnamefont {Moessner}},\ and\
  \bibinfo {author} {\bibfnamefont {P.~A.}\ \bibnamefont {McClarty}},\
  }\bibfield  {title} {\bibinfo {title} {Magnon interactions in the frustrated
  pyrochlore ferromagnet
  ${\mathrm{yb}}_{2}{\mathrm{ti}}_{2}{\mathrm{o}}_{7}$},\ }\href
  {https://doi.org/10.1103/PhysRevB.100.104423} {\bibfield  {journal} {\bibinfo
   {journal} {Phys. Rev. B}\ }\textbf {\bibinfo {volume} {100}},\ \bibinfo
  {pages} {104423} (\bibinfo {year} {2019})}\BibitemShut {NoStop}%
\bibitem [{\citenamefont {Hickey}\ \emph {et~al.}(2025)\citenamefont {Hickey},
  \citenamefont {Lozano-G\'omez},\ and\ \citenamefont
  {Gingras}}]{Hickey_et_al_2025_PRB}%
  \BibitemOpen
  \bibfield  {author} {\bibinfo {author} {\bibfnamefont {A.}~\bibnamefont
  {Hickey}}, \bibinfo {author} {\bibfnamefont {D.}~\bibnamefont
  {Lozano-G\'omez}},\ and\ \bibinfo {author} {\bibfnamefont {M.~J.~P.}\
  \bibnamefont {Gingras}},\ }\bibfield  {title} {\bibinfo {title}
  {Order-by-disorder without quantum zero-point fluctuations in the pyrochlore
  heisenberg ferromagnet with dzyaloshinskii-moriya interactions},\ }\href
  {https://doi.org/10.1103/PhysRevB.111.184434} {\bibfield  {journal} {\bibinfo
   {journal} {Phys. Rev. B}\ }\textbf {\bibinfo {volume} {111}},\ \bibinfo
  {pages} {184434} (\bibinfo {year} {2025})}\BibitemShut {NoStop}%
\bibitem [{\citenamefont {Li}\ and\ \citenamefont
  {Chen}(2018)}]{Li_Chen_spin_1_pyrochlore_2018}%
  \BibitemOpen
  \bibfield  {author} {\bibinfo {author} {\bibfnamefont {F.-Y.}\ \bibnamefont
  {Li}}\ and\ \bibinfo {author} {\bibfnamefont {G.}~\bibnamefont {Chen}},\
  }\bibfield  {title} {\bibinfo {title} {Competing phases and topological
  excitations of spin-1 pyrochlore antiferromagnets},\ }\href
  {https://doi.org/10.1103/PhysRevB.98.045109} {\bibfield  {journal} {\bibinfo
  {journal} {Phys. Rev. B}\ }\textbf {\bibinfo {volume} {98}},\ \bibinfo
  {pages} {045109} (\bibinfo {year} {2018})}\BibitemShut {NoStop}%
\bibitem [{\citenamefont {Laurell}\ and\ \citenamefont
  {Fiete}(2017)}]{Laurell_Fiete_2017_prl}%
  \BibitemOpen
  \bibfield  {author} {\bibinfo {author} {\bibfnamefont {P.}~\bibnamefont
  {Laurell}}\ and\ \bibinfo {author} {\bibfnamefont {G.~A.}\ \bibnamefont
  {Fiete}},\ }\bibfield  {title} {\bibinfo {title} {Topological magnon bands
  and unconventional superconductivity in pyrochlore iridate thin films},\
  }\href {https://doi.org/10.1103/PhysRevLett.118.177201} {\bibfield  {journal}
  {\bibinfo  {journal} {Phys. Rev. Lett.}\ }\textbf {\bibinfo {volume} {118}},\
  \bibinfo {pages} {177201} (\bibinfo {year} {2017})}\BibitemShut {NoStop}%
\bibitem [{\citenamefont {Jian}\ and\ \citenamefont
  {Nie}(2018)}]{Jian_Nie_Weyl_2018}%
  \BibitemOpen
  \bibfield  {author} {\bibinfo {author} {\bibfnamefont {S.-K.}\ \bibnamefont
  {Jian}}\ and\ \bibinfo {author} {\bibfnamefont {W.}~\bibnamefont {Nie}},\
  }\bibfield  {title} {\bibinfo {title} {Weyl magnons in pyrochlore
  antiferromagnets with an all-in-all-out order},\ }\href
  {https://doi.org/10.1103/PhysRevB.97.115162} {\bibfield  {journal} {\bibinfo
  {journal} {Phys. Rev. B}\ }\textbf {\bibinfo {volume} {97}},\ \bibinfo
  {pages} {115162} (\bibinfo {year} {2018})}\BibitemShut {NoStop}%
\bibitem [{\citenamefont {Hwang}\ \emph {et~al.}(2020)\citenamefont {Hwang},
  \citenamefont {Trivedi},\ and\ \citenamefont
  {Randeria}}]{hwang_et_al_prl_2020}%
  \BibitemOpen
  \bibfield  {author} {\bibinfo {author} {\bibfnamefont {K.}~\bibnamefont
  {Hwang}}, \bibinfo {author} {\bibfnamefont {N.}~\bibnamefont {Trivedi}},\
  and\ \bibinfo {author} {\bibfnamefont {M.}~\bibnamefont {Randeria}},\
  }\bibfield  {title} {\bibinfo {title} {Topological magnons with nodal-line
  and triple-point degeneracies: Implications for thermal hall effect in
  pyrochlore iridates},\ }\href
  {https://doi.org/10.1103/PhysRevLett.125.047203} {\bibfield  {journal}
  {\bibinfo  {journal} {Phys. Rev. Lett.}\ }\textbf {\bibinfo {volume} {125}},\
  \bibinfo {pages} {047203} (\bibinfo {year} {2020})}\BibitemShut {NoStop}%
\bibitem [{\citenamefont {Jyothis}\ \emph {et~al.}(2024)\citenamefont
  {Jyothis}, \citenamefont {Patra},\ and\ \citenamefont
  {Chandra}}]{Jyothis_et_al_JPCM_2024}%
  \BibitemOpen
  \bibfield  {author} {\bibinfo {author} {\bibfnamefont {V.~V.}\ \bibnamefont
  {Jyothis}}, \bibinfo {author} {\bibfnamefont {B.}~\bibnamefont {Patra}},\
  and\ \bibinfo {author} {\bibfnamefont {V.~R.}\ \bibnamefont {Chandra}},\
  }\bibfield  {title} {\bibinfo {title} {Magnon bands in pyrochlore slabs with
  heisenberg exchange and anisotropies},\ }\href
  {https://doi.org/10.1088/1361-648X/ad21aa} {\bibfield  {journal} {\bibinfo
  {journal} {Journal of Physics: Condensed Matter}\ }\textbf {\bibinfo {volume}
  {36}},\ \bibinfo {pages} {185801} (\bibinfo {year} {2024})}\BibitemShut
  {NoStop}%
\bibitem [{\citenamefont {Dzyaloshinsky}(1958)}]{Dzyaloshinsky_1958}%
  \BibitemOpen
  \bibfield  {author} {\bibinfo {author} {\bibfnamefont {I.}~\bibnamefont
  {Dzyaloshinsky}},\ }\bibfield  {title} {\bibinfo {title} {A thermodynamic
  theory of “weak” ferromagnetism of antiferromagnetics},\ }\href
  {https://doi.org/https://doi.org/10.1016/0022-3697(58)90076-3} {\bibfield
  {journal} {\bibinfo  {journal} {Journal of Physics and Chemistry of Solids}\
  }\textbf {\bibinfo {volume} {4}},\ \bibinfo {pages} {241} (\bibinfo {year}
  {1958})}\BibitemShut {NoStop}%
\bibitem [{\citenamefont {Moriya}(1960{\natexlab{a}})}]{Moriya_DMI_prl}%
  \BibitemOpen
  \bibfield  {author} {\bibinfo {author} {\bibfnamefont {T.}~\bibnamefont
  {Moriya}},\ }\bibfield  {title} {\bibinfo {title} {New mechanism of
  anisotropic superexchange interaction},\ }\href
  {https://doi.org/10.1103/PhysRevLett.4.228} {\bibfield  {journal} {\bibinfo
  {journal} {Phys. Rev. Lett.}\ }\textbf {\bibinfo {volume} {4}},\ \bibinfo
  {pages} {228} (\bibinfo {year} {1960}{\natexlab{a}})}\BibitemShut {NoStop}%
\bibitem [{\citenamefont {Moriya}(1960{\natexlab{b}})}]{Moriya_DMI_Phys_Rev}%
  \BibitemOpen
  \bibfield  {author} {\bibinfo {author} {\bibfnamefont {T.}~\bibnamefont
  {Moriya}},\ }\bibfield  {title} {\bibinfo {title} {Anisotropic superexchange
  interaction and weak ferromagnetism},\ }\href
  {https://doi.org/10.1103/PhysRev.120.91} {\bibfield  {journal} {\bibinfo
  {journal} {Phys. Rev.}\ }\textbf {\bibinfo {volume} {120}},\ \bibinfo {pages}
  {91} (\bibinfo {year} {1960}{\natexlab{b}})}\BibitemShut {NoStop}%
\bibitem [{\citenamefont {Moessner}\ and\ \citenamefont
  {Chalker}(1998{\natexlab{a}})}]{moessner_and_chalker_PRL_1998}%
  \BibitemOpen
  \bibfield  {author} {\bibinfo {author} {\bibfnamefont {R.}~\bibnamefont
  {Moessner}}\ and\ \bibinfo {author} {\bibfnamefont {J.~T.}\ \bibnamefont
  {Chalker}},\ }\bibfield  {title} {\bibinfo {title} {Properties of a classical
  spin liquid: The heisenberg pyrochlore antiferromagnet},\ }\href
  {https://doi.org/10.1103/PhysRevLett.80.2929} {\bibfield  {journal} {\bibinfo
   {journal} {Phys. Rev. Lett.}\ }\textbf {\bibinfo {volume} {80}},\ \bibinfo
  {pages} {2929} (\bibinfo {year} {1998}{\natexlab{a}})}\BibitemShut {NoStop}%
\bibitem [{\citenamefont {Moessner}\ and\ \citenamefont
  {Chalker}(1998{\natexlab{b}})}]{moessner_and_chalker_PRB_1998}%
  \BibitemOpen
  \bibfield  {author} {\bibinfo {author} {\bibfnamefont {R.}~\bibnamefont
  {Moessner}}\ and\ \bibinfo {author} {\bibfnamefont {J.~T.}\ \bibnamefont
  {Chalker}},\ }\bibfield  {title} {\bibinfo {title} {Low-temperature
  properties of classical geometrically frustrated antiferromagnets},\ }\href
  {https://doi.org/10.1103/PhysRevB.58.12049} {\bibfield  {journal} {\bibinfo
  {journal} {Phys. Rev. B}\ }\textbf {\bibinfo {volume} {58}},\ \bibinfo
  {pages} {12049} (\bibinfo {year} {1998}{\natexlab{b}})}\BibitemShut {NoStop}%
\bibitem [{\citenamefont {Elhajal}\ \emph {et~al.}(2005)\citenamefont
  {Elhajal}, \citenamefont {Canals}, \citenamefont {Sunyer},\ and\
  \citenamefont {Lacroix}}]{elhajal_et_al_pyrochlore_dm}%
  \BibitemOpen
  \bibfield  {author} {\bibinfo {author} {\bibfnamefont {M.}~\bibnamefont
  {Elhajal}}, \bibinfo {author} {\bibfnamefont {B.}~\bibnamefont {Canals}},
  \bibinfo {author} {\bibfnamefont {R.}~\bibnamefont {Sunyer}},\ and\ \bibinfo
  {author} {\bibfnamefont {C.}~\bibnamefont {Lacroix}},\ }\bibfield  {title}
  {\bibinfo {title} {Ordering in the pyrochlore antiferromagnet due to
  dzyaloshinsky-moriya interactions},\ }\href
  {https://doi.org/10.1103/PhysRevB.71.094420} {\bibfield  {journal} {\bibinfo
  {journal} {Phys. Rev. B}\ }\textbf {\bibinfo {volume} {71}},\ \bibinfo
  {pages} {094420} (\bibinfo {year} {2005})}\BibitemShut {NoStop}%
\bibitem [{\citenamefont {Luttinger}\ and\ \citenamefont
  {Tisza}(1946)}]{luttinger_and_tisza_1946}%
  \BibitemOpen
  \bibfield  {author} {\bibinfo {author} {\bibfnamefont {J.~M.}\ \bibnamefont
  {Luttinger}}\ and\ \bibinfo {author} {\bibfnamefont {L.}~\bibnamefont
  {Tisza}},\ }\bibfield  {title} {\bibinfo {title} {Theory of dipole
  interaction in crystals},\ }\href {https://doi.org/10.1103/PhysRev.70.954}
  {\bibfield  {journal} {\bibinfo  {journal} {Phys. Rev.}\ }\textbf {\bibinfo
  {volume} {70}},\ \bibinfo {pages} {954} (\bibinfo {year} {1946})}\BibitemShut
  {NoStop}%
\bibitem [{\citenamefont {Holstein}\ and\ \citenamefont
  {Primakoff}(1940)}]{holstein_primakoff_original_paper_1940}%
  \BibitemOpen
  \bibfield  {author} {\bibinfo {author} {\bibfnamefont {T.}~\bibnamefont
  {Holstein}}\ and\ \bibinfo {author} {\bibfnamefont {H.}~\bibnamefont
  {Primakoff}},\ }\bibfield  {title} {\bibinfo {title} {Field dependence of the
  intrinsic domain magnetization of a ferromagnet},\ }\href
  {https://doi.org/10.1103/PhysRev.58.1098} {\bibfield  {journal} {\bibinfo
  {journal} {Phys. Rev.}\ }\textbf {\bibinfo {volume} {58}},\ \bibinfo {pages}
  {1098} (\bibinfo {year} {1940})}\BibitemShut {NoStop}%
\bibitem [{\citenamefont {Colpa}(1978)}]{colpa_diagonalisation_1978}%
  \BibitemOpen
  \bibfield  {author} {\bibinfo {author} {\bibfnamefont {J.}~\bibnamefont
  {Colpa}},\ }\bibfield  {title} {\bibinfo {title} {Diagonalization of the
  quadratic boson hamiltonian},\ }\href
  {https://doi.org/https://doi.org/10.1016/0378-4371(78)90160-7} {\bibfield
  {journal} {\bibinfo  {journal} {Physica A: Statistical Mechanics and its
  Applications}\ }\textbf {\bibinfo {volume} {93}},\ \bibinfo {pages} {327 }
  (\bibinfo {year} {1978})}\BibitemShut {NoStop}%
\bibitem [{\citenamefont {Fetter}\ and\ \citenamefont
  {Walecka}(1971)}]{Fetter_and_Walecka}%
  \BibitemOpen
  \bibfield  {author} {\bibinfo {author} {\bibfnamefont {A.~L.}\ \bibnamefont
  {Fetter}}\ and\ \bibinfo {author} {\bibfnamefont {J.~D.}\ \bibnamefont
  {Walecka}},\ }\href@noop {} {\emph {\bibinfo {title} {Quantum theory of
  many-particle systems}}}\ (\bibinfo  {publisher} {New York (N.Y.) :
  McGraw-Hill},\ \bibinfo {year} {1971})\BibitemShut {NoStop}%
\bibitem [{\citenamefont {Wick}(1950)}]{Wicks_Theorem_original_paper}%
  \BibitemOpen
  \bibfield  {author} {\bibinfo {author} {\bibfnamefont {G.~C.}\ \bibnamefont
  {Wick}},\ }\bibfield  {title} {\bibinfo {title} {The evaluation of the
  collision matrix},\ }\href {https://doi.org/10.1103/PhysRev.80.268}
  {\bibfield  {journal} {\bibinfo  {journal} {Phys. Rev.}\ }\textbf {\bibinfo
  {volume} {80}},\ \bibinfo {pages} {268} (\bibinfo {year} {1950})}\BibitemShut
  {NoStop}%
\bibitem [{\citenamefont {Harris}\ \emph {et~al.}(1992)\citenamefont {Harris},
  \citenamefont {Kallin},\ and\ \citenamefont
  {Berlinsky}}]{Harris_Kallin_Berlinsky_PRB_1992}%
  \BibitemOpen
  \bibfield  {author} {\bibinfo {author} {\bibfnamefont {A.~B.}\ \bibnamefont
  {Harris}}, \bibinfo {author} {\bibfnamefont {C.}~\bibnamefont {Kallin}},\
  and\ \bibinfo {author} {\bibfnamefont {A.~J.}\ \bibnamefont {Berlinsky}},\
  }\bibfield  {title} {\bibinfo {title} {Possible n\'eel orderings of the
  kagom\'e antiferromagnet},\ }\href {https://doi.org/10.1103/PhysRevB.45.2899}
  {\bibfield  {journal} {\bibinfo  {journal} {Phys. Rev. B}\ }\textbf {\bibinfo
  {volume} {45}},\ \bibinfo {pages} {2899} (\bibinfo {year}
  {1992})}\BibitemShut {NoStop}%
\bibitem [{\citenamefont {Starykh}\ \emph {et~al.}(2006)\citenamefont
  {Starykh}, \citenamefont {Chubukov},\ and\ \citenamefont
  {Abanov}}]{Starykh_Chubukov_Abanov_PRB_2006}%
  \BibitemOpen
  \bibfield  {author} {\bibinfo {author} {\bibfnamefont {O.~A.}\ \bibnamefont
  {Starykh}}, \bibinfo {author} {\bibfnamefont {A.~V.}\ \bibnamefont
  {Chubukov}},\ and\ \bibinfo {author} {\bibfnamefont {A.~G.}\ \bibnamefont
  {Abanov}},\ }\bibfield  {title} {\bibinfo {title} {Flat spin-wave dispersion
  in a triangular antiferromagnet},\ }\href
  {https://doi.org/10.1103/PhysRevB.74.180403} {\bibfield  {journal} {\bibinfo
  {journal} {Phys. Rev. B}\ }\textbf {\bibinfo {volume} {74}},\ \bibinfo
  {pages} {180403(R)} (\bibinfo {year} {2006})}\BibitemShut {NoStop}%
\bibitem [{\citenamefont {Maksimov}\ and\ \citenamefont
  {Chernyshev}(2016)}]{Chernyshev_Maksimov_Honeycomb_PRB_2016}%
  \BibitemOpen
  \bibfield  {author} {\bibinfo {author} {\bibfnamefont {P.~A.}\ \bibnamefont
  {Maksimov}}\ and\ \bibinfo {author} {\bibfnamefont {A.~L.}\ \bibnamefont
  {Chernyshev}},\ }\bibfield  {title} {\bibinfo {title} {Field-induced
  dynamical properties of the $\mathit{XXZ}$ model on a honeycomb lattice},\
  }\href {https://doi.org/10.1103/PhysRevB.93.014418} {\bibfield  {journal}
  {\bibinfo  {journal} {Phys. Rev. B}\ }\textbf {\bibinfo {volume} {93}},\
  \bibinfo {pages} {014418} (\bibinfo {year} {2016})}\BibitemShut {NoStop}%
\bibitem [{\citenamefont {Zhitomirsky}\ and\ \citenamefont
  {Nikuni}(1998)}]{Zhitomirsky_and_Nikuni_prb_1998}%
  \BibitemOpen
  \bibfield  {author} {\bibinfo {author} {\bibfnamefont {M.~E.}\ \bibnamefont
  {Zhitomirsky}}\ and\ \bibinfo {author} {\bibfnamefont {T.}~\bibnamefont
  {Nikuni}},\ }\bibfield  {title} {\bibinfo {title} {Magnetization curve of a
  square-lattice heisenberg antiferromagnet},\ }\href
  {https://doi.org/10.1103/PhysRevB.57.5013} {\bibfield  {journal} {\bibinfo
  {journal} {Phys. Rev. B}\ }\textbf {\bibinfo {volume} {57}},\ \bibinfo
  {pages} {5013} (\bibinfo {year} {1998})}\BibitemShut {NoStop}%
\bibitem [{\citenamefont {Zhitomirsky}\ and\ \citenamefont
  {Chernyshev}(1999)}]{Zhitomirsky_and_Chernyshev_sq_lat_PRL_1999}%
  \BibitemOpen
  \bibfield  {author} {\bibinfo {author} {\bibfnamefont {M.~E.}\ \bibnamefont
  {Zhitomirsky}}\ and\ \bibinfo {author} {\bibfnamefont {A.~L.}\ \bibnamefont
  {Chernyshev}},\ }\bibfield  {title} {\bibinfo {title} {Instability of
  antiferromagnetic magnons in strong fields},\ }\href
  {https://doi.org/10.1103/PhysRevLett.82.4536} {\bibfield  {journal} {\bibinfo
   {journal} {Phys. Rev. Lett.}\ }\textbf {\bibinfo {volume} {82}},\ \bibinfo
  {pages} {4536} (\bibinfo {year} {1999})}\BibitemShut {NoStop}%
\end{thebibliography}%
\end{document}